\documentclass[aps,prl,preprint,groupedaddress,showkeys]{revtex4-1}

\usepackage{graphicx}
\usepackage{dcolumn}
\usepackage{bm}
\usepackage{braket}
\usepackage{color}
\newcommand{\red}[1]{\textcolor{black}{#1}}
\begin{document}


\title{ Multi-Dimensional self-exciting NBD  process \\
and  
\\  Default portfolios
}

\author{Masato Hisakado}
\email{hisakadom@yahoo.co.jp} 
\affiliation{
 Nomura Holdings, Inc., Otemachi 2-2-2, Chiyoda-ku, Tokyo 100-8130, Japan}. 
\author{Kodai Hattori}
\email{shintaro.mori@gmail.com}
\author{Shintaro Mori}
\email{shintaro.mori@gmail.com}
\affiliation{
\dag Department of Mathematics and Physics,
Graduate School of Science and Technology, 
Hirosaki University \\
Bunkyo-cho 3, Hirosaki, Aomori 036-8561, Japan}

\date{\today}
\begin{abstract} 
In this study,  we apply  a multidimensional
self-exciting negative binomial distribution (SE-NBD) process
 to default portfolios with 13 sectors.
 The SE-NBD process is a Poisson process with a gamma-distributed intensity function.
 We extend the SE-NBD process to a multidimensional process.
 Using the multidimensional SE-NBD process (MD-SE-NBD),
 we can  estimate interactions between these 13 sectors
 as a network.
 By applying impact analysis, we can classify upstream and downstream sectors.
 The upstream sectors are real-estate and financial institution (FI) sectors.
 From these upstream sectors, shock spreads to the downstream sectors.
 This is an amplifier of the shock.
 This is consistent with the analysis of bubble bursts.
 We compare these results to the multidimensional Hawkes process (MD-Hawkes) that
  has a zero-variance intensity function.

\hspace{0cm}
\vspace{1cm}
\end{abstract}

\maketitle
\bibliography{basename of .bib file}
\newpage
\section{I. Introduction}

In financial engineering, several products have been developed to hedge against risks.
A credit default swap is a tool used to hedge credit risks.
These products protect against a subset of the total loss on a credit portfolio in exchange for payments, and
provide valuable insights into market implications regarding default dependencies such as clustering of defaults \cite{Hisakado,M2010,M2008,Sch}. 
This aspect is important because the difficulties in managing credit events depend on these correlations.
To represent the clustering defaults as a time series, 
the Hawkes process was recently applied \cite{Haw,kir,qp,err}.
It is a Poisson process with a  zero-variance intensity function \cite{KZ, KZ1}.
In the Hawkes process with a conditional intensity function, as the number of events increases, probability of the events increases as well.
This corresponds to self-excitation and causes temporal correlation.
This process exhibits a phase transition 
between steady and non-steady states.

In our previous study,
we introduced the self-exciting negative binomial (SE-NBD) process that has a correlation in same term and temporal correlation \cite{Hisakado10}.
The correlation in the same term is the same as that considered in the P\'{o}lya urn model, while the temporal correlation is incorporated by introducing the conditional initial condition.
The SE-NBD process is a Poisson process with a gamma-distributed intensity function.
In the SE-NBD process with a conditional intensity function, as the number of events increases, probability of the events also increases, similarly to the Hawkes process with the conditional intensity function.

The SE-NBD process is within the double-scaling limit of the urn process \cite{Hisakado10}. 
In the standard continuous limit, such an urn process becomes the Hawkes process that has no correlations in the same term.
Both processes have the same phase transition point from the steady state to the non-steady state.
The difference between these processes is variance in the intensity function.
The variance in the intensity function of the SE-NBD is non-zero, that is, gamma distribution.

For the default portfolio, the results produced by the urn process are superior to those obtained using the Hawkes process \cite{Hisakado10}.
This is because of the network effects that correspond to the variance of the intensity function.
 In fact, some firms affect many other firms, while some others do not affect other firms.
This is among the network effects because 
 a network with hubs has a large variance in degree distributions \cite{Hisakado7}.
In this study, we used  13 sector portfolios to extend this research.


The remainder of this paper is organized as follows.
In section II, we review the SE-NBD process.
In section III, we introduce the multidimensional SE-NBD process.
In section IV, 
we discuss continuous limits of the process and correlation functions.
In section V,
we describe application of this process to the default history empirical
data and discuss the network. 
Finally, we draw conclusions in section VI.

\section{II. Review of negative binomial distribution process}

In this section, we review the discrete self-exciting negative binomial distribution (SE-NBD) process \cite{Hisakado10}.
$X_t$ denotes the number of events at time $t$ and 
$\hat{d}_i$ denotes the discount factor for the temporal correlation and kernel function.
Moreover, we considered exponential decay $\hat{d}_i=r^{i-1}$ in this research.
This corresponds to a short memory \cite{Long}.
The number of events at time $t$ corresponds to negative binomial distribution (NBD),
\begin{eqnarray}
P(X_{t+1}&=&k_{t+1})\sim \mbox{NBD}(X_{t+1}=k_{t+1}|K_t, M_t/K_t=M_0/K_0) \nonumber\\
&=&\frac{(K_t+k_{t+1}-1)!}{k_{t+1}!(K_t-1)!}
(\frac{K_t}{K_t+M_t})^{K_t}
(\frac{M_t}{K_t+M_t})^{k_{t+1}},
\end{eqnarray}
where 
\begin{equation}
M_{t}
=M_0+M_0/L_0\sum_i^{t}k_i \hat{d}_{t+1-i},
\label{m}
\end{equation}
and
\begin{equation}
K_{t}=
K_0+K_0/L_0\sum_i^{t}k_i\hat{d}_{t+1-i},
\end{equation}
and 
\begin{equation}
M_{t}/K_t=M_0/K_0.
\end{equation}
The conditional probability is defined by updating the parameters $K_t$ and $M_t$.
Note that for  all process parameters $M_t/K_t$, $M_0/K_0$ is constant,
 which simultaneously corresponds to the correlation at the same time.
Under this condition, the process has the reproductive properties of an NBD.
$L_0$ and $\hat{d}_i$ are considered as the parameters for the temporal correlation. 


The negative binomial distribution $\mbox{NBD}(X_{t+1}=k_{t+1}|K_t,M_t/K_t=M_0/K_0)$ has the following form:
\begin{eqnarray}
\mbox{NBD}(X_{t+1}=k_{t+1}|K_t,M_0/K_0)&=&
\int_0^{\infty}
\mbox{Poisson}(k_{t+1}|\lambda_{t+1}) \cdot \mbox{Gamma}(\lambda_{t+1} |K_t,M_0/K_0) d \lambda_{t+1},
\nonumber \\
&=&
\int_0^{\infty} 
\frac{\lambda^{k_{t+1}}e^{-\lambda_{t+1}}}{k_{t+1} !}\dot
\frac{\lambda_{t+1}^{K_t-1}}{\Gamma(K_t)(M_0/K_0)^{K_t}}e^{-\lambda K_0/M_0} d\lambda_{t+1},
\label{lambda}
\end{eqnarray}
where $\mbox{Poisson}(k_{t+1}|\lambda_{t+1})$ is a Poisson process and
$\mbox{Gamma}(\lambda_{t+1} |K_t,M_0/K_0)$ is a gamma distribution.
The intensity function $\lambda_{t+1}$ follows a gamma distribution.
Note that $\mbox{Gamma} (\lambda |K_t,M_0/K_0)$ has average $M_t$ and variance $M_t^2/K_t$.
The intensity function has variance compared to the
Poisson process, which has a zero-variance intensity function, namely, the Hawkes process \cite{Haw}.
We call this the discrete self-exciting negative binomial distribution (SE-NBD) process.
Self-excitation is introduced by conditional probability through $K_t$ and $M_t$.
When $M_t=M_0$ and $K_t=K_0$, the non-self-exciting case is referred to as the NBD process.

In summary, we
obtain a discrete SE-NBD process
$X_{t}$ that obeys NBD for $M_t$ as,
\begin{equation}
X_{t+1}\sim \mbox{NBD}
\left(K_{t},M_0/K_0 \right),t\ge 0,
\label{SENBD}
\end{equation}
where 
\begin{equation}
M_t=M_0+M_0/L_0\sum_{s=1}^{t}X_s \hat{d}_{t+1-s},t\ge 1, 
\label{SE1}
\end{equation}
and 
\begin{equation}
K_t=K_0+K_0/L_0\sum_{s=1}^{t}X_s \hat{d}_{t+1-s},t\ge 1.  
\end{equation}
In the limit $K_0 \rightarrow \infty$, we obtain a discrete Hawkes process \cite{kir}.
$X_{t}$ follows a Poisson process for $M_t$,
\begin{equation}
X_{t+1}\sim \mbox{Poisson}
\left(M_{t} \right),t\ge 0, 
\label{Hawkes}
\end{equation}
where 
\begin{equation}
M_t=M_0+M_0/L_0\sum_{s=1}^{t}X_s \hat{d}_{t+1-s},t\ge 1.  
\end{equation}

Here, we set the average $\bar{v}$ of the intensity function.
The mean-field approximation in Eq.(\ref{SE1}) is as follows:
\begin{equation}
\bar{v}=M_0/(1-(M_0/L_0) \hat{T}),
\label{mfa}
\end{equation}
 where $\hat{T}=\sum^{\infty} \hat{d}_{i}=1/(1-r)$.
 When $M_0/L_0\hat{T}<1$, we obtain a steady state.
 However, when $M_0/L_0\hat{T}\geq 1$,
 the non-steady state is observed.
 This denotes the phase transition between the steady and non-steady
 states.
 This self-consistent equation and transition point are the same as those for the discrete Hawkes process, as described in Eq.(\ref{mfa}).

\section{III. Multi Dimensional SE-NBD process}

\subsection{A. Introduction of Multi Dimensional SE-NBD process}
We extended the discrete SE-NBD process to a multidimensional model (MD-SE-NBD).  
$X_{t}^{(i)}$ obeys NBD for  $M_t^{(i)}$, where $i=1,2,\cdots, D$, 
\begin{equation}
X_{t+1}^{(i)}\sim \mbox{NBD}
\left(K_{t}^{(i)},M_0^{(i)}/K_0^{(i)} \right),t\ge 0, 
\end{equation}
where 
\begin{equation}
M_t^{(i)}=M_0^{(i)}+\sum_{j=1}^{D}M_0^{(i)}/L_0^{(ij)}\sum_{s=1}^{t}X_s^{(j)} \hat{d}_{t+1-s}^{(i)},t\ge 1,
\label{M}
\end{equation}
and 
\begin{equation}
K_t^{(i)}=K_0^{(i)}+\sum_{j=1}^{D}K_0^{(i)}/L_0^{(ij)}\sum_{s=1}^{t}X_s^{(j)} \hat{d}_{t+1-s}^{(i)},t\ge 1,  
\end{equation}
where $\hat{d}_s^{(i)}=r_i^{s-1}$ denotes the exponential decay.
In this model, the interaction from $i$ to $j$ is $M_0^{(j)}/L_0^{(ji)}$
and from $j$ to $i$ is $M_0^{(i)}/L_0^{(ij)}$
.
Note that, in this model, the interaction is asymmetrical.

Accordingly, we also consider the discrete multidimensional Hawkes process (MD-Hawkes)
$X_{t}^{(i)}$ that follows a Poisson distribution for $M_t^{(i)}$, where $i=1,2,\cdots, D$,
\begin{equation}
X_{t+1}^{(i)}\sim \mbox{Poisson}
\left(M_{t}^{(i)} \right),t\ge 0, 
\end{equation}
and where 
\begin{equation}
M_t^{(i)}=M_0^{(i)}+\sum_{j=1}^{D}M_0^{(i)}/L_0^{(ij)}\sum_{s=1}^{t}X_s^{(j)} \hat{d}_{t+1-s}^{(i)},t\ge 1,
\end{equation}
and additionally, where $\hat{d}_s^{(i)}=r_i^{s-1}$ denotes the exponential decay.

Here, we consider the interaction between lines $j$ and $i$
as $S_{ij}$,
\begin{equation}
  S_{ij}=(M_0^{(i)}/L_0^{(ij)})T^{(i)},  
  \label{S}
\end{equation}
where $T^{(i)}=\sum_{s=1}^{\infty} \hat{d}_{s}^{(i)}$.
In the matrix form, we can obtain the following expression:
\begin{equation}
\mathbf{S}\equiv\left(
    \begin{array}{cccc}
     S_{11} &  S_{12}  &\cdots& S_{1D} \\
   S_{21}      &  S_{22} & \ddots&\vdots\\
 \vdots & \ddots & \ddots  & \vdots\\
    S_{D1}&  \cdots &   \cdots & S_{DD}  \\
    \end{array}
  \right).
  \label{matrix}
\end{equation}
\red{
In a single case, we considered an event.
The Total expected number of events affected by an event
is 
\[
\frac{M_0}{L_0}\sum_{s=1}^\infty \hat{d}_{s}.
\label{im2}
\]
In the case $\hat{d}_i=r^{i-1}$, it becomes 
$M_0/L_0 (1-r)$.
$S_{ij}$ is the multidimensional extension of this formula and denotes the number of affected events in the $i$-th line by an event in the $j$-th line.
This is the direct effect of the event.
In the next subsection, we discuss the indirect effects of an event through the network.
}
That is, parameter $S_{ij}$ is the effective reproduction number
regarding an infectious disease from lines $j$ to $i$.
This denotes the number of patients infected by one patient 
in the infection model.
Hence, we can study the interaction between lines $i$ and
$j$ using matrix $\mathbf{S}$.

Here, we set the average $\bar{v}^{(i)}$ of the $i$-th line and vector $\mathbf{\bar{v}}=\bar{v}^{(i)}$;
then, we obtain the matrix-form equations for $\bar{v}^{(i)}$ using Eq.(\ref{M}),
\begin{equation}
\mathbf{\bar{v}}=\mathbf{M_0}+\mathbf{S}\mathbf{\bar{v}},
\label{M3}
\end{equation}
where \red{$\mathbf{M_0}$ is the vector $\mathbf{M_0}=M_0^{(i)}$. Solving Eq.(\ref{M3}), we obtain the equilibrium solution for $\mathbf{\bar{v}}$,
\begin{equation}
\mathbf{\bar{v}}=(\mathbf{E}-\mathbf{S})^{-1}\mathbf{M_0},
\label{mfa2}
\end{equation}
}
where $\mathbf{E}$ is the unit matrix.
Eq.(\ref{mfa2}) corresponds to Eq.(\ref{mfa}) in the single-line case, while Eq.(\ref{M3}) denotes the mean-field approximation of the 
MD-SE-NBD process.
We can confirm the transition point between the steady and non-steady states, that is, $\mathbf{E}=\mathbf{S}$ described in Eq.(\ref{mfa2}) for a multidimensional case. 
Note that, in the same way, we can obtain the equilibrium solution for the hybrid case \red{that} includes both the SE-NBD and Hawkes processes.
This means that some lines are MD-SE-NBD and the rest are MD-Hawkes processes that interact with each other.

\red{\subsection{B. Impact analysis of MD-SE-NBD process}
In this subsection
 we consider the sensitivity against impact.
 In other words, we study the increase in future events when  an event is added through the network.
 This denotes the sensitivity to adding an event.
 }
To study the sensitivity against impact, we considered 
the first shock for the $i$-th line at time $0$.
This is to compare sectors using a network such as Google's page rank \cite{page}.
The shock vector is  $\mathbf{v}_0^{(i)}=(0,\cdots,1,\cdots,0)$, in which only the
$i$-th element is $1$ and others are $0$.
This corresponds to an event on the $i$-th line.
 We considered the increase in events due to this impact.
In the single-line case, the added number of events for the shock event is
\begin{equation}
    v_{\infty}=\frac{M_0}{L_0}\sum_{i=1}^{\infty}(\frac{M_0}{L_0}+r)^i =\frac{M_0}{L_0}\frac{1}{1-r-M_0/L_0},
    \label{im}
\end{equation}
where  $\hat{d}_i=r^{i-1}$.
\red{The proof is stated in Appendix A.}
Note that $M_0/L_0=1-r$ is the transition point.
We extend this to a multidimensional case.

\begin{figure}[h]
\includegraphics[width=110mm]{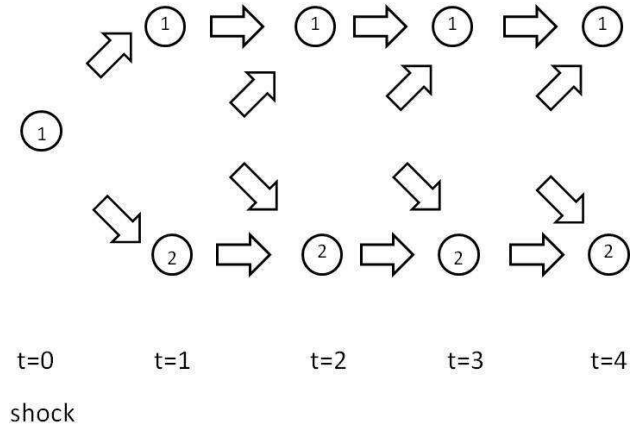}
\caption{Contagion of shock for the two-dimensional case. The impact decays exponentially.}
\label{shock}
\end{figure}

Here, we introduce two matrices $\mathbf{\hat{S}}$ 
\begin{equation}
  \mathbf{\hat{S}}=\hat{S}^{(ij)}=M_0^{(i)}/L_0^{(ij)},  
\end{equation}  
and 
the diagonal matrix, $\mathbf{\hat{T}}$, 
\begin{equation}
  \mathbf{\hat{T}}=\hat{T}^{(ii)}=r_{(i)}.  
\end{equation}

Here, we consider the parameter vector
\begin{equation}
 \mathbf{M}_t=(M^{(i)}_t).   
\end{equation}
$\mathbf{M}_t^+$ includes effects of the impact  $\mathbf{v}_0^{(i)}$, while  $\mathbf{M}_t$ does not include the impact.
The effect of the impact at time $t$, $\mathbf{v}_t^{(i)}$, is defined as the difference 
\begin{equation}
 \mathbf{v}_t^{(i)}=\sum_{s=1}^{t}\mbox{E}[M_s^{(i)+}-M_s^{(i)}],   
\end{equation}
where \red{$E(x)$} denotes the expected value of $x$.
This is the sum of the increases in the expected value of the intensity function for shock at time $0$.
In other words, it signifies the expected number of events for 
shock in the future.
 We can obtain $\mathbf{v}_t^{(i)}$, 
 
\begin{equation}
\mathbf{v}_t^{(i)}=\mathbf{\hat{S}}\sum_{j=1}^{t}(\mathbf{\hat{T}}+\mathbf{\hat{S}})^j \mathbf{v}_0^{(i)},
\end{equation}
where $j=t$ is the first contagion and $j=0, \cdots, t-1$ denotes the temporal decay of the impacts.

In the limit $t\rightarrow \infty$, we calculate
$\mathbf{v}_{\infty}^{(i)}$,
\begin{equation}
 \mathbf{v}_{\infty}^{(i)}=\mathbf{\hat{S}}\sum_{j=1}^{\infty} (\mathbf{\hat{T}}+\mathbf{\hat{S}})^j \mathbf{v}_{0}^{(i)} =
 (\mathbf{E}-\mathbf{S})^{-1} \mathbf{{S}}\mathbf{v}_{0}^{(i)}.
 \label{imp}
\end{equation}
\red{It is an extension of Eq.(\ref{im}).}
$\mathbf{v}_{\infty}^{(i)}$ denotes the increase in the number of events in each line for the shock of $\mathbf{v}_{0}^{(i)}$.
\red{The sum of the elements of the vector
$\mathbf{v}_{\infty}=\sum_i \mathbf{v}^{(i)}_{\infty}/D$ denotes the total   increase in the number of
events for the average shock, that is, $\sum_i \mathbf{v}_{0}^{(i)}/D$.
}
The shock of the upstream region has a larger impact than that of the downstream region.
We discuss the relationship between the impact analysis and branching process in Appendix B.
Note that we can confirm the transition point $\mathbf{E}=\mathbf{S}$ depicted in Eq.(\ref{imp}).

\section{VI. Continuous limit}

In this section, we discuss the continuous limit of the 
discrete MD-SE-NBD model.
Here, we introduce the counting process $\tilde{X}_t^{(i)}=\sum_{s} X_s^{(i)}$.
$M_{t}^{(i)}$ is the average of the $i$-th line, as defined in Eq.(\ref{M}).
We set $M_{t}^{(i)}=\theta_t^{(i)}\Delta$.
 We can confirm that $M_{t}^{(i)}$ increases  as $\Delta$ increases, and $\Delta$ plays the role of time.
 Moreover, we set $M_{t}^{(i)}/K_{t}^{(i)}=\omega_i'$, which is independent of $t$.
 Hence, the process has the reproductive properties of an NBD.
 
 We obtain the mean of the intensity function in the continuous limit at $t$, $\lambda_t^{(i)}$,
 which is the intensity function at $t$, as
depicted in Eq.(\ref{lambda}) for the single-line case,
\begin{equation}
E(\lambda_t^{(i)}|F_t)=\lim_{\Delta\rightarrow 0}\frac{E[\tilde{X}_{t+\Delta}^{(i)}-\tilde{X}_t^{(i)}|F_t]}{\Delta}=\lim_{\Delta\rightarrow 0}\frac{M_t^{(i)}}{\Delta}=(\theta_0^{(i)}+\sum_{j=1}^{D}\sum_{s<t} \tilde{\omega}_{ij}X_s^{(j)}\hat{d}_{t-s}^{(i)} ),
\label{av}
\end{equation}
where $F_t$ is the history of the number of events and 
$\tilde{\omega}_{ij}=\theta^{(i)}/L_0^{(ij)}$.
Eq.(\ref{av}) corresponds to the hazard function \cite{Sch}.

The variance in the intensity of the distribution at time $t$ is 
\begin{equation}
Var(\lambda_t^{(i)}|F_t)=\lim_{\Delta\rightarrow 0}\frac{(M_t^{(i)})^2/K_t^{(i)}}{\Delta}=\omega'_{i}   ( \theta_0^{(i)}+\sum_{j=1}^{D}\sum_{s<t} \tilde{\omega}_{ij}\hat{d}_{t-s}^{(i)} X_s^{(i)})\red{.}
\label{ba}
\end{equation}

In the continuous MD-SE-NBD process, the intensity function follows a gamma distribution as a discrete MD-SE-NBD process.
In summary, we can obtain, in the continuous limit, 
\begin{equation}
\tilde{X}_{t+\Delta}^{(i)}-\tilde{X}_{t}^{(i)}\sim \mbox{NBD}
\left(\theta_t^{(i)}\Delta/\omega'_i,\omega'_i\right),t\ge 0, 
\label{av2}
\end{equation}
where 
\begin{equation}
\theta_t^{(i)}=\theta_0^{(i)}+\sum_{j=1}^{D} \sum_{s<t} \tilde{\omega}_{ij}X_s^{(j)}\hat{d}_{t-s}^{(i)},t\ge 0.  \label{SE}
\end{equation}

In the limit $\omega'_{i}\rightarrow 0$, the variance of the intensity function vanishes. Subsequently, the continuous SE-NBD process becomes the Hawkes process.
\begin{equation}
\tilde{X}^{(i)}_{t+\Delta}-\tilde{X}^{(i)}_{t}\sim \mbox{Poisson}
\left(\theta_t^{(i)}\Delta \right),t\ge 0, 
\label{av3}
\end{equation}
where 
\begin{equation}
\theta_t^{(i)}=\theta_0^{(i)}+\sum_{j=1}^{D} \sum_{s<t} \tilde{\omega}_{ij}X_s^{(j)}\hat{d}_{t-s}^{(i)},t\ge 0.  \label{HP}
\end{equation}

Next, we consider the correlation functions of the continuous MD-SE-NBD process.
The calculations are presented in detail in Appendix C.
We can obtain the covariance as,
\begin{equation}
    E[X_t^{(i)},X_{t+\tau}^{(j)}]=[C_{ij}(\tau)+\bar{v}_i\bar{v}_j+(1+\omega'_i)\bar{v}_i\delta_{ij}\delta(\tau)]\Delta^2,
\end{equation}
where $C_{ij}(\tau)$ denotes the covariance density function for $\tau>0$, $\delta_{ij}$ denotes the Kronecker's delta, and $\delta(x)$ denotes the delta function.
There is no correlation between the $i$-th line and $j$-th line \red{at} the same time.
The SE-NBD process has a larger correlation at the same time than the Hawkes process.

Here, we define $g_{ij}(x)=\tilde{\omega}_{ij}\hat{d}_x^{(i)}$ and set $\tau>0$.
We can obtain the integral equation for the correlation function 
using Eq.(\ref{mfa2})
\begin{equation}
    C_{ik}(\tau)=(\omega'_i+1)\bar{v}_i g_{ii}(\tau)+\sum_{j=1}^{D}\int_{0}^{\infty}
    g_{kj}(w)C_{ij}(\tau-w)dw.
    \label{ieq2}
\end{equation}
 The MD-Hawkes process is the case where $\omega'_i=0$.  
The difference between the SE-NBD and Hawkes processes is the correlation in the same term, which is the first term on the RHS.

\section{V. Parameter estimation for the default data }

In this section, we present the application of parameter estimation to quarterly default data.
S\&P default data from 1981 to 2020 were used in this study. 
The data included 13 sectors \cite{Data1}.
We classified the 13 sectors as follows: 1) forest and building products/home builders, 2) consumer/service sector, 3) energy and natural resources, 4) financial institutions (FI), 5) healthcare/chemical, 6) hitech/computers/office equipment, 7) insurance, 8) leisure time/media, 9) aerospace/automotive/capital goods/metal, 10) real estate, 11) telecommunications, 12) transportation, and 13) utility.
We considered the interaction as a network among the aforementioned sectors.

We estimated the parameters using the Bayes’ formula:
\begin{eqnarray}
P(K_0, M_0,L_0, r| k_1, \cdots, k_T)
&=&\frac{P(k_T|\red{K_0,M_0,L_0,} r))}{P(k_T)}
\cdots
\frac{P(k_1|K_0,M_0,L_0,r)}{P(k_1)}
\nonumber \\
&\times&f(K_0,M_0,L_0,r),
\label{MAP}
\end{eqnarray}
where $f(K_0,M_0,L_0,r)$ is the prior distribution \cite{Hisakado6}.
We used a uniform distribution for the prior distribution.
We applied the maximum a-posteriori (MAP) estimation described in Eq.(\ref{MAP}).
If we use the NBD for distribution $P$,
the process is the parameter estimation for the discrete MD-SE-NBD process introduced in Section III.
If we use the Poisson distribution instead of the NBD, the process is the parameter estimation for the discrete MD-Hawkes process.
In this case, the parameters were $M_0$, $L_0$, and $r$.

We applied the following five models: (1) multidimensional SE-NBD model (MD-SE-NBD), (2) multidimensional Hawkes model (MD-Hawkes), (3) SE-NBD model, (4) Hawkes model, and (5) NBD model.
Models  (3), (4), and (5) do not include interactions among sectors.
Model (5) does not include self-excitation.
Note that we can fit the model independently as sector models because of the definition of models (1) MD--SE-NBD and (2) MD-Hawkes.
The models are summarized in Table.\ref{models}.
The AIC values of the models are reported in Table.\ref{AIC}.
The parameters for models (1), (2), (3), (4), and (5) are listed in Tables.\ref{MDSENBD}, \ref{MDH}, and \ref{Fig4}.
Here, $S_{ij}$ denotes the interaction among sectors, 
defined in Eq.(\ref{S}).

\begin{table}[tbh]
\caption{The summary  of the models }
\begin{center}
\begin{tabular}{|l|l|c|c|c|c|c|}
\multicolumn{4}{c}{}\\ \hline
No.& p
Property  &(1) MD-SE-NBD & (2)  MD-Hawkes  & (3)  SE-NBD& (4) Hawkes & (5) NBD \\
 \hline \hline
1&Variance of  intensity function& $\bigcirc$ & & $\bigcirc$&  & $\bigcirc$\\ \hline
2&Interaction among sectors& $\bigcirc$ &$\bigcirc$ &&  & \\ \hline
3& Self exciting & $\bigcirc$ & $\bigcirc$ &$\bigcirc$&$\bigcirc$&\\ \hline
\end{tabular}
\label{models}
\end{center}
\end{table}

From Table.\ref{AIC},
we can confirm that (1) MD-SE-NBD has the lowest AIC among all the sectors \red{except} real estate.
For the real-estate sector (2) MD-Hawkes has the lowest AIC.
In fact, we can confirm the large $K_0$ of the real-estate sector with models (1) MD-SE-NBD and (3) SE-NBD from 
Table.\ref{MDSENBD} and \ref{Fig4}.
A large $K_0$ means that the variance of the intensity function is small, similarly to the Hawkes model.
The Hawkes model has a zero-\red{variance} intensity function.
In other words,
\red{except} the real-estate sector, these processes have a
large variance in the intensity function.

Comparing the AIC of (1) MD-SE-NBD and (3) SE-NBD,
we can confirm the effect of the interaction owing to other sectors based on 
the smaller AIC of (1) MD-SE-NBD in all the sectors. 
In (3) SE-NBD,
there is no interaction among the sectors.
This can also be confirmed by comparing the AIC of the
(2) MD-Hawkes and 
(4) Hawkes models because the 
MD-Hawkes model has a smaller AIC.


Comparing the AIC of (3) SE-NBD and (5) NBD,
we can confirm the effect of self-excitation because 
\red{of} the smaller AIC of (3) SE-NBD in all the sectors.
In (5) NBD, there is no self-excitation.

\begin{figure}[h]
\begin{center}
\begin{tabular}{c}
\begin{minipage}{0.5\hsize}
\begin{center}
\includegraphics[clip, width=8cm]{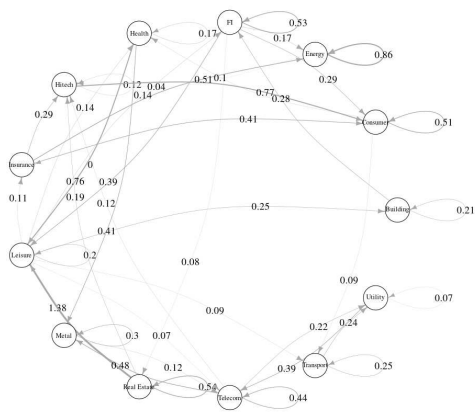}
\hspace{1.6cm} (a)
\end{center}
\end{minipage}
\begin{minipage}{0.33\hsize}
\begin{center}
\includegraphics[clip, width=8cm]{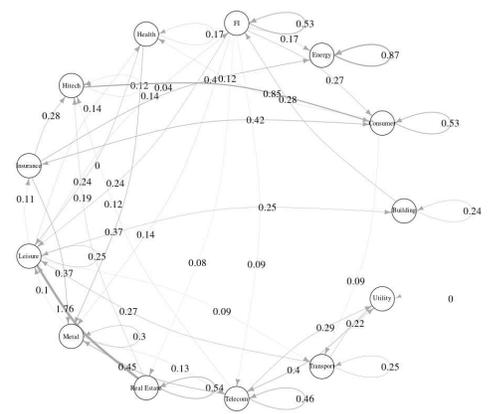}
 \hspace{1.6cm} (b)
\end{center}
\end{minipage}

 \end{tabular}
\caption{Interaction among the sectors: (a)  MD-SE-NBD and (b) MD-Hawkes models}
\label{correlation}
\end{center}
\end{figure}

Next, we confirm the interaction among the sectors using
(1) MD-SE-NBD and (2) MD-NBD models.
The interactions are represented by 
matrix $S$, defined in Eq.(\ref{S}).
 The interactions between the sectors is depicted in Figs. \ref{correlation} (a) and (b) using the parameters defined in Table.\ref{MDSENBD}, Table.\ref{MDH}, and Table.\ref{Fig4}.
In the case of the Hawkes process, these interactions are called Hawkes graphs \cite{Emb}.
However,  it is difficult to confirm the interactions between upstream and downstream
sectors of these networks.

\begin{figure}[htbp]
\begin{tabular}{ccc}    
\includegraphics[width=5.5cm]{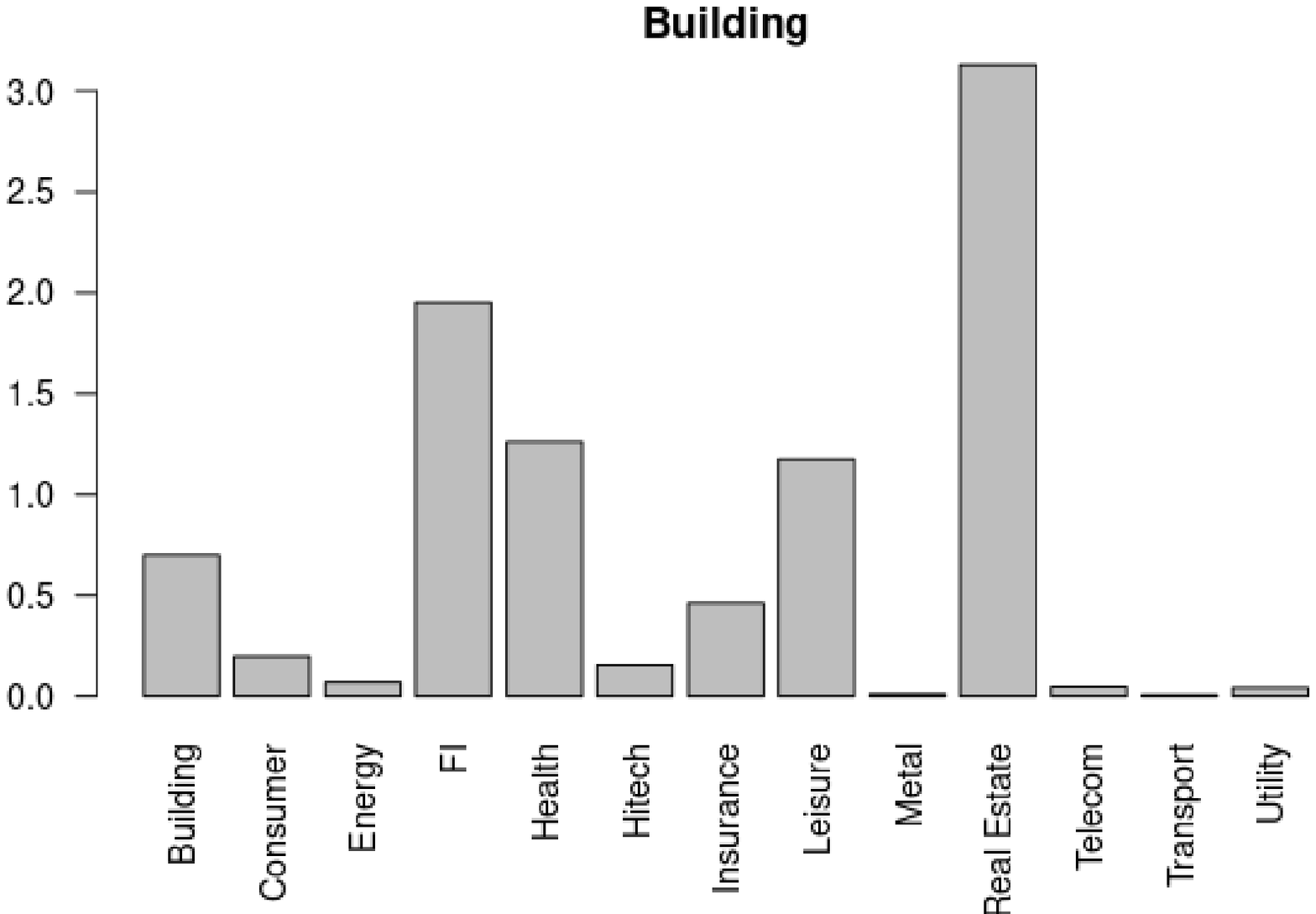} &
\includegraphics[width=5.5cm]{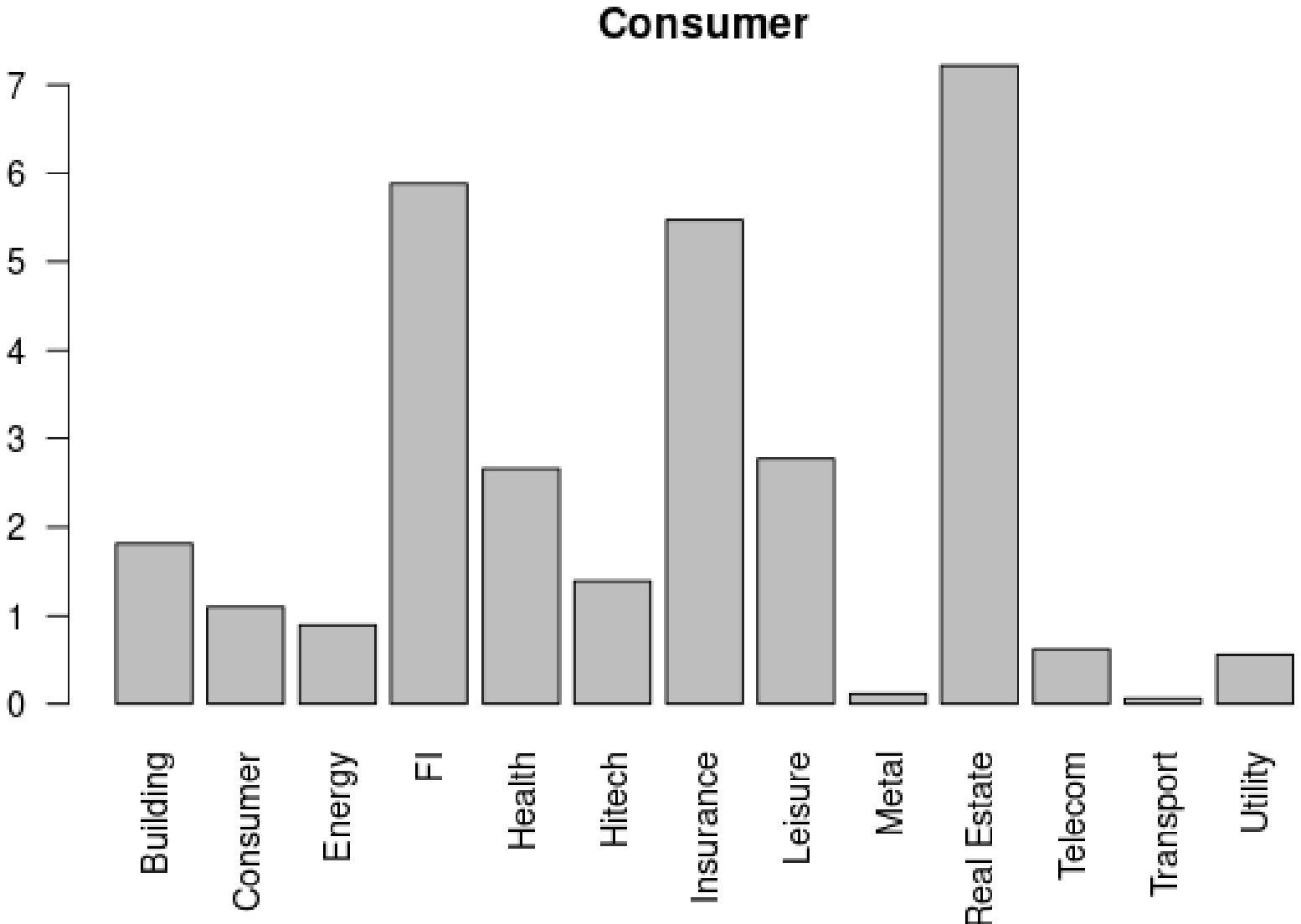} &
\includegraphics[width=5.5cm]{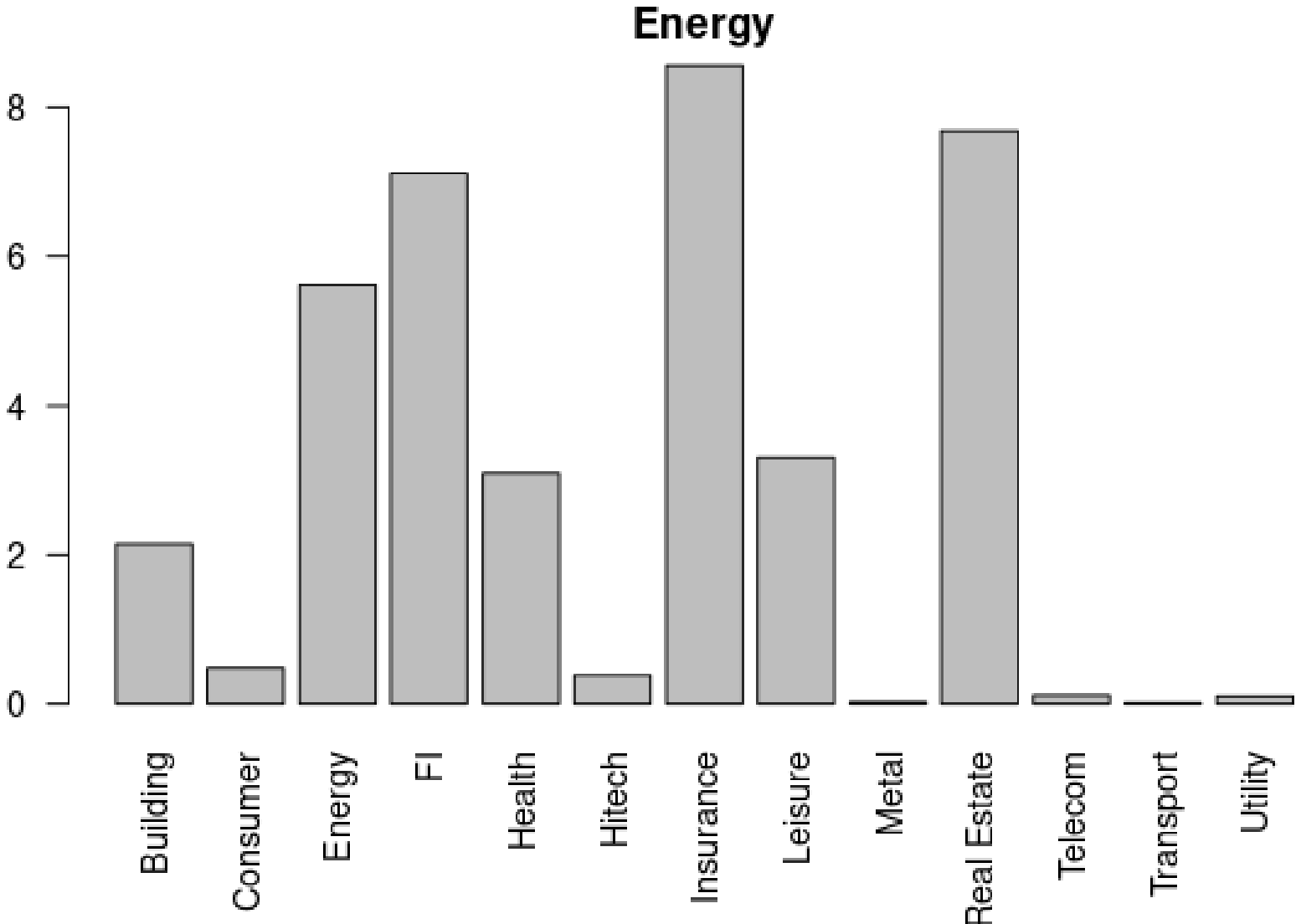} \\
\includegraphics[width=5.5cm]{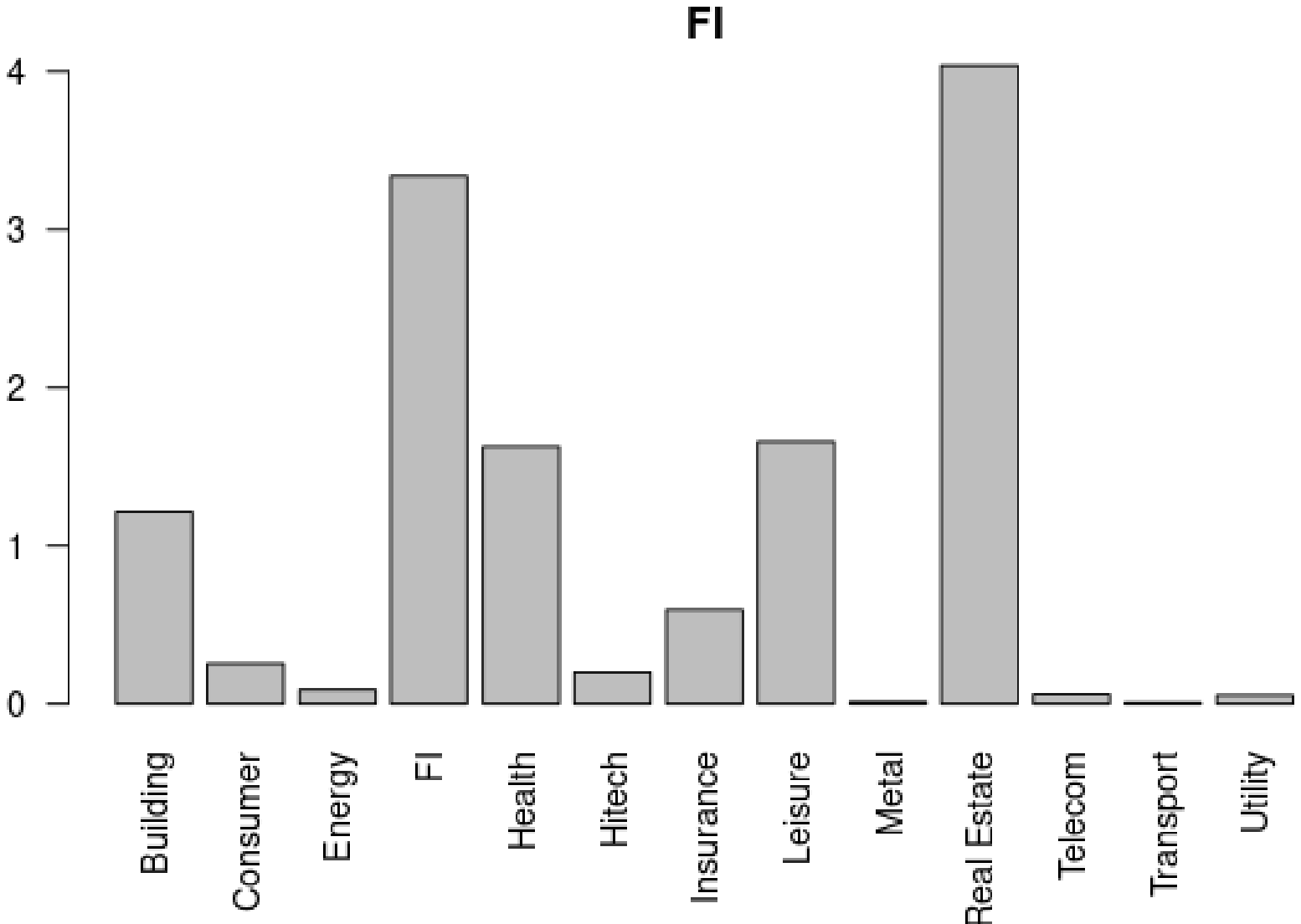} &
\includegraphics[width=5.5cm]{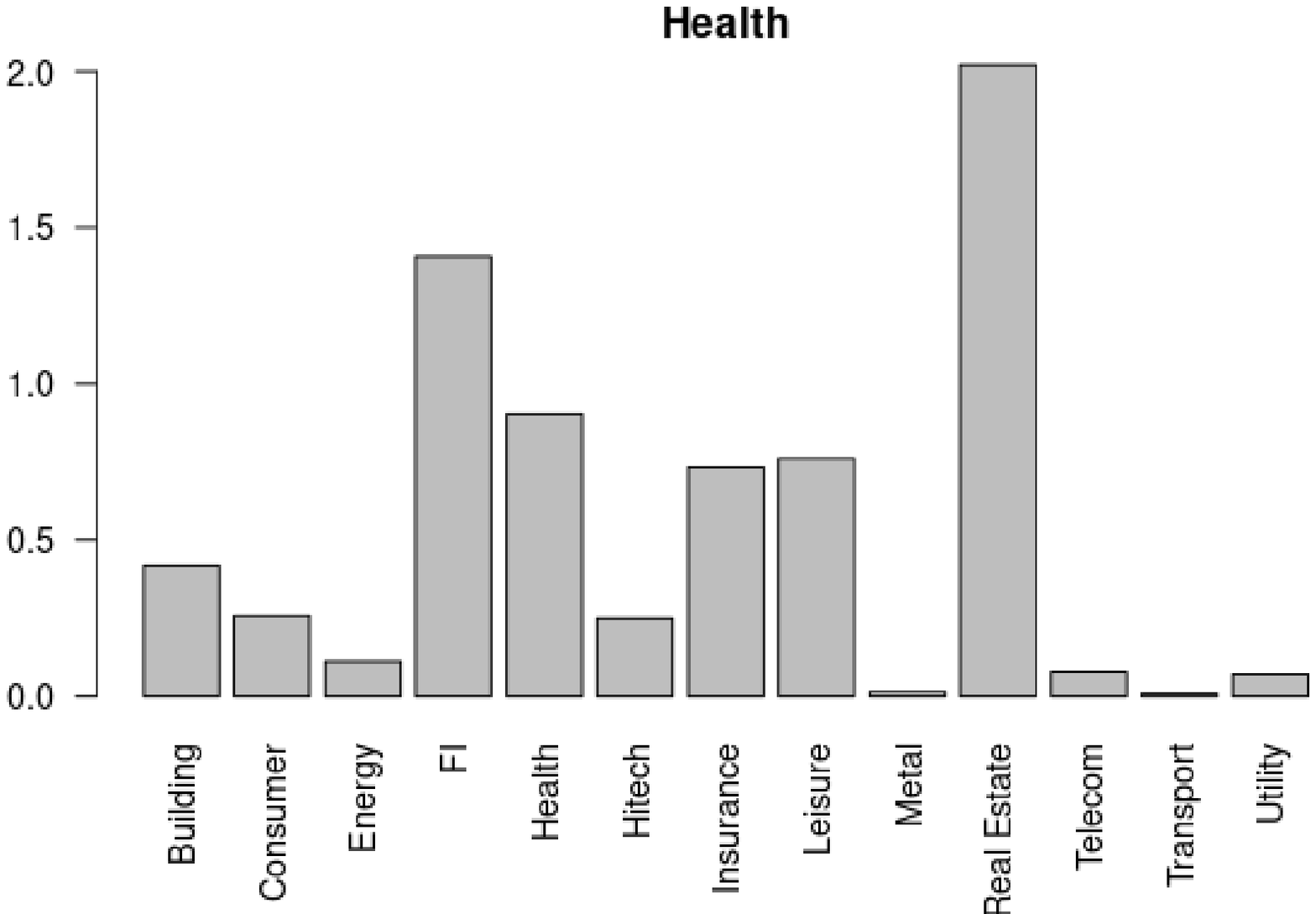} &
\includegraphics[width=5.5cm]{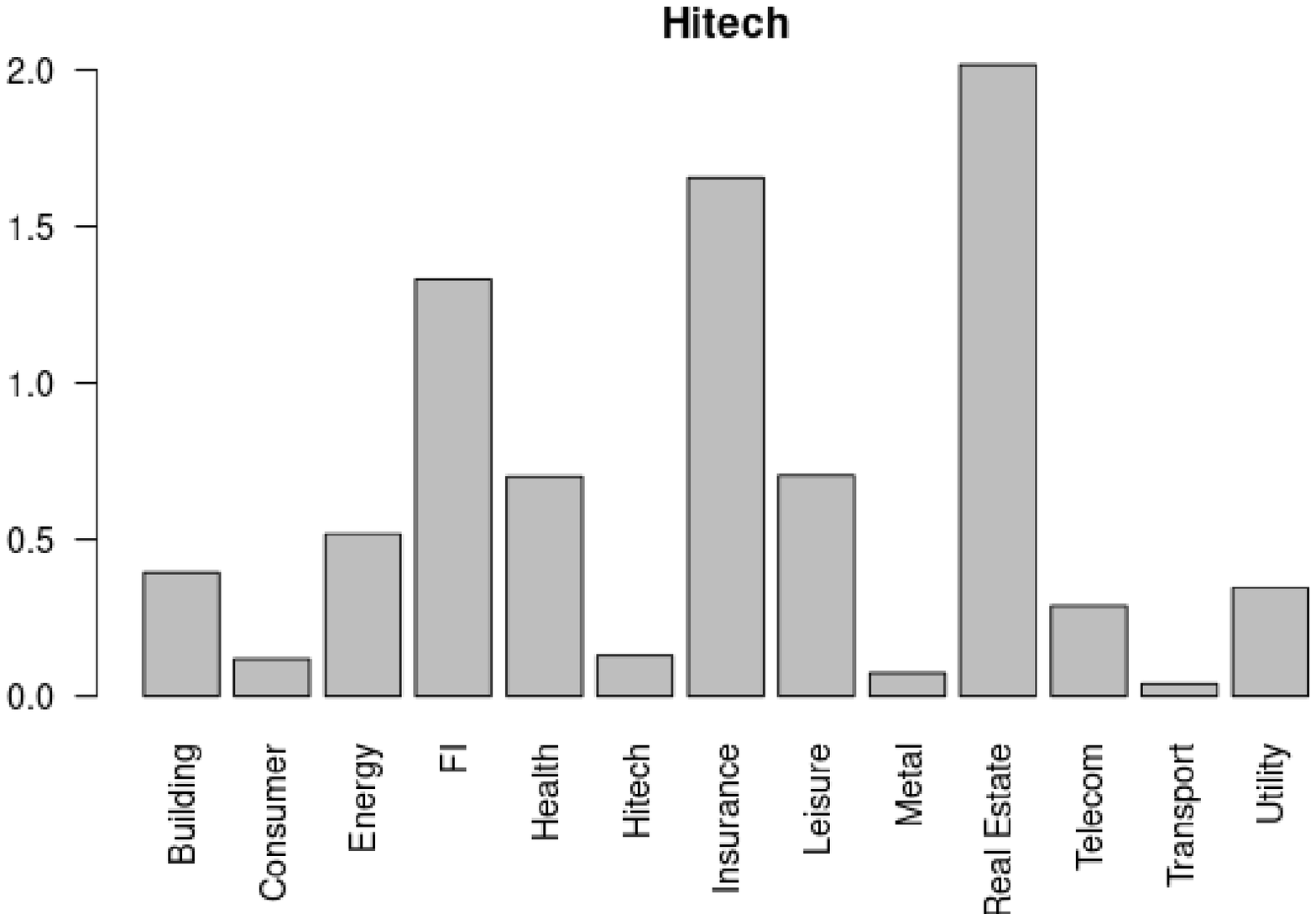} \\
\includegraphics[width=5.5cm]{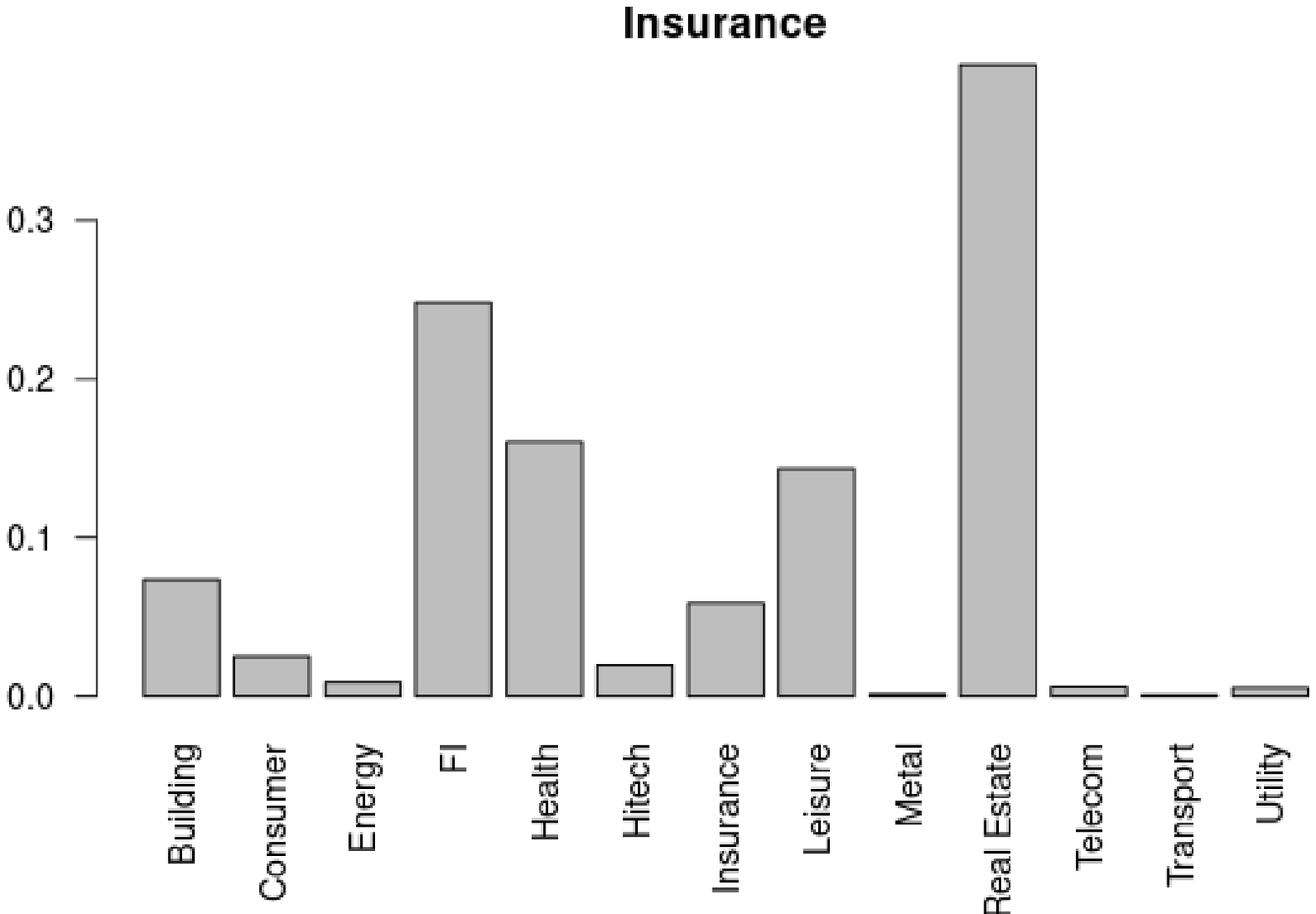} &
\includegraphics[width=5.5cm]{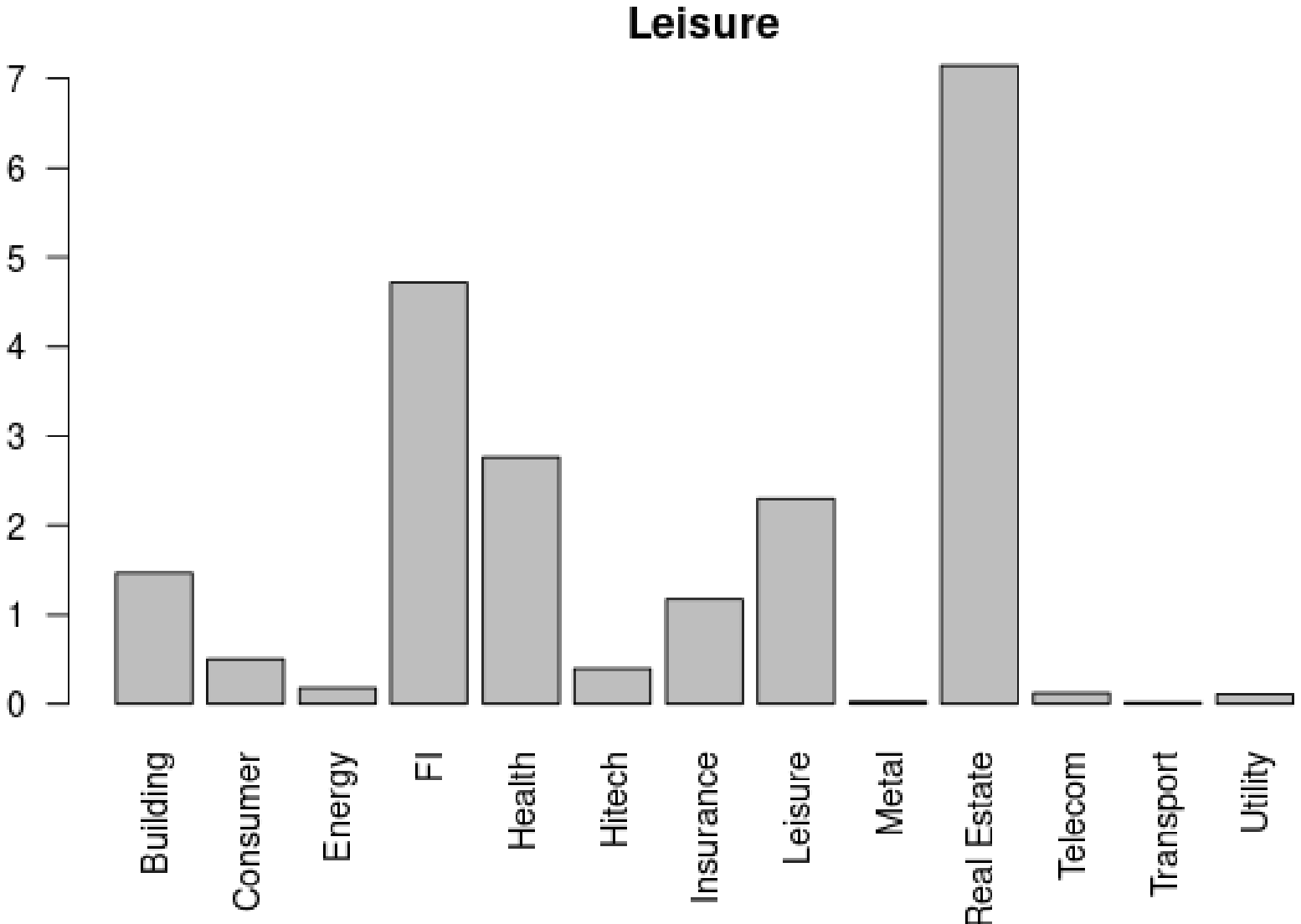} &
\includegraphics[width=5.5cm]{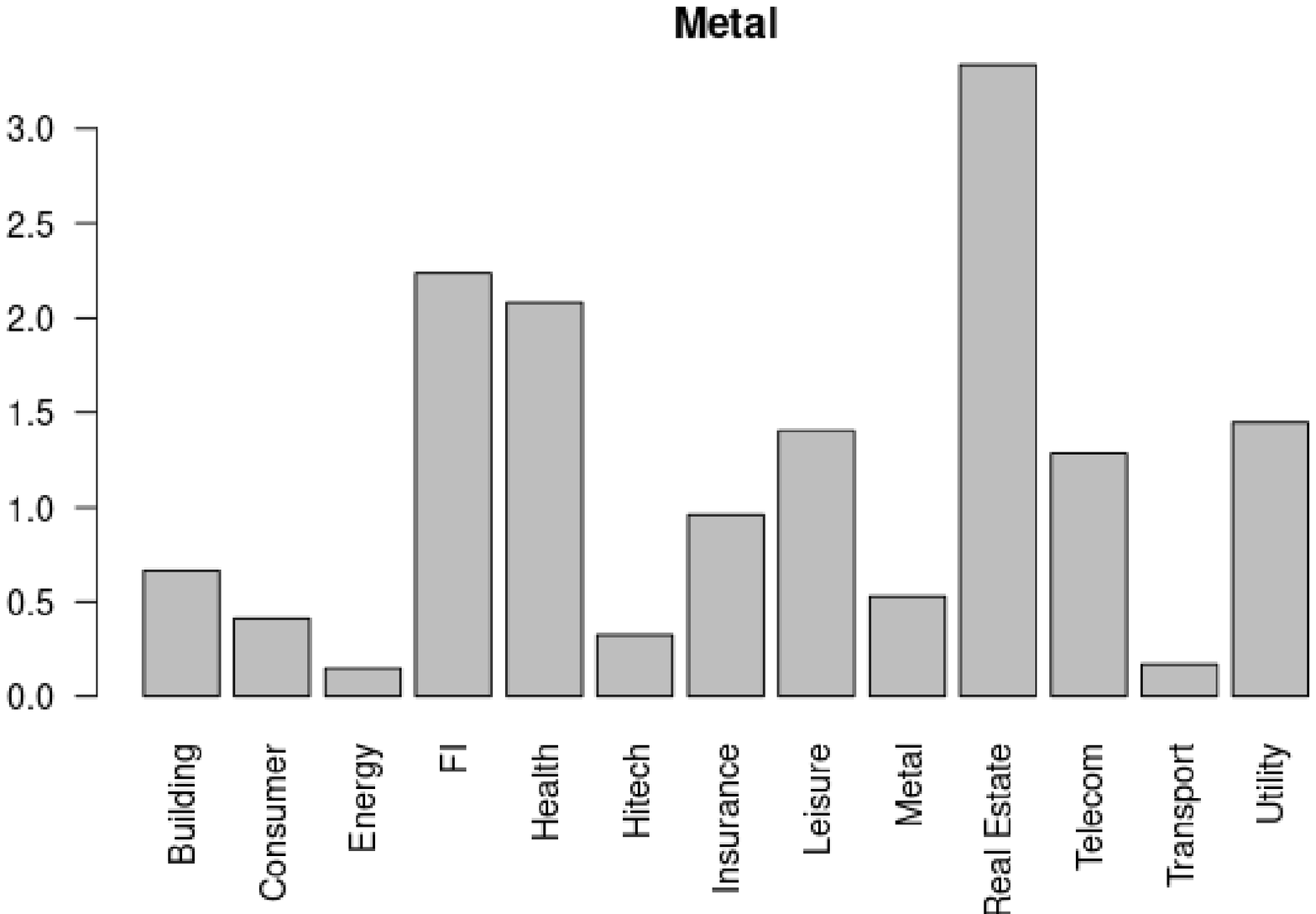} \\
\includegraphics[width=5.5cm]{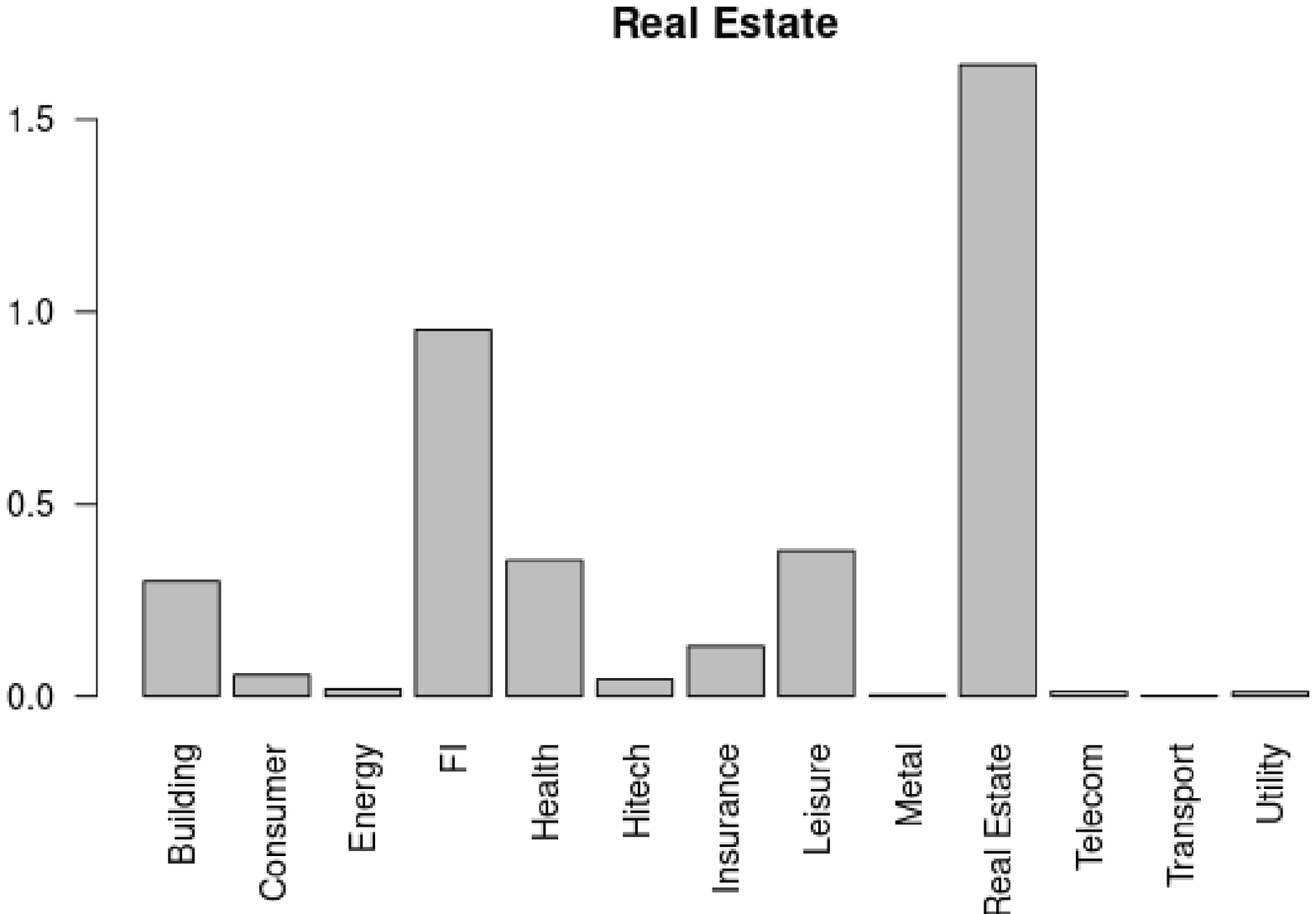} &
\includegraphics[width=5.5cm]{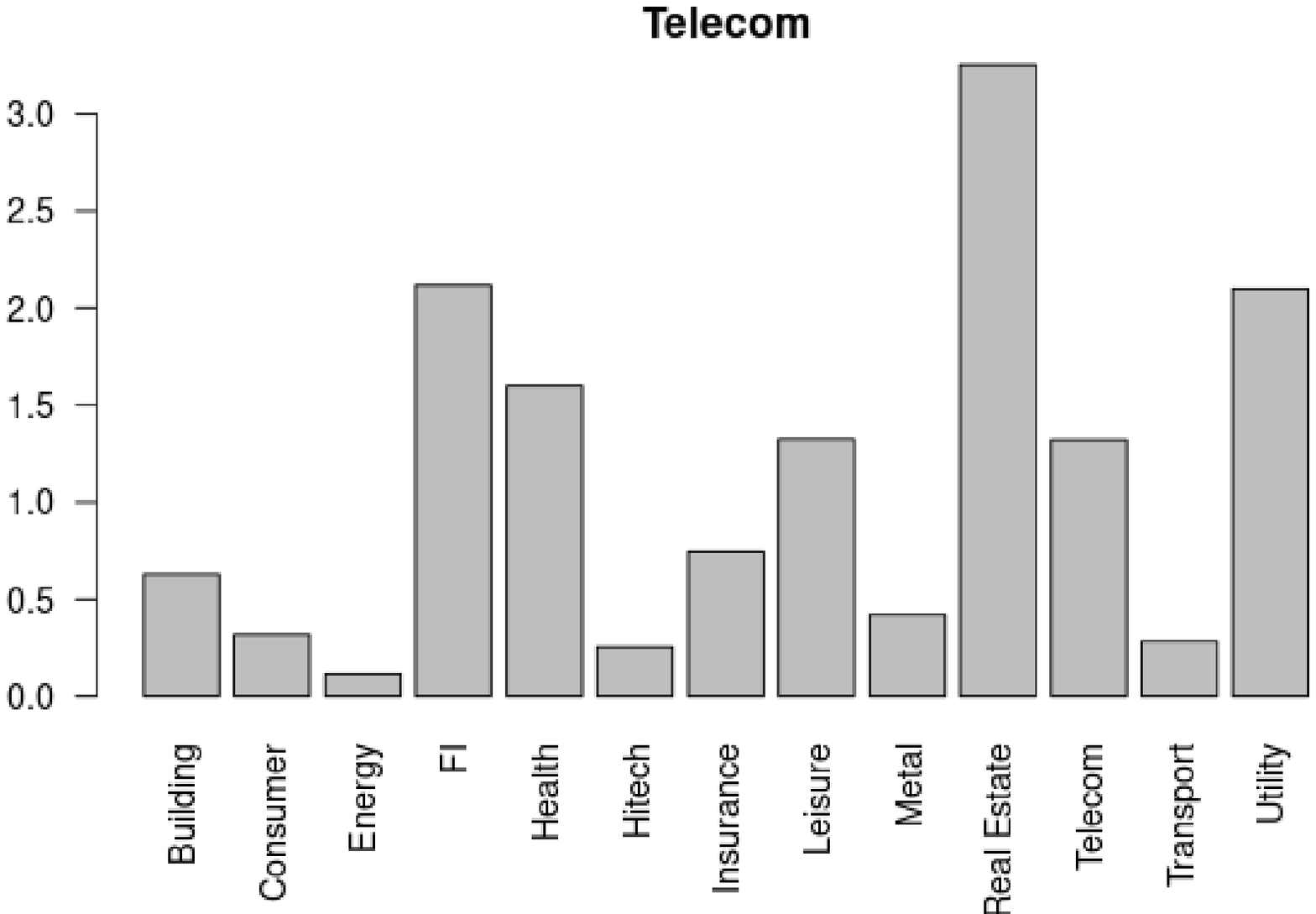} &
\includegraphics[width=5.5cm]{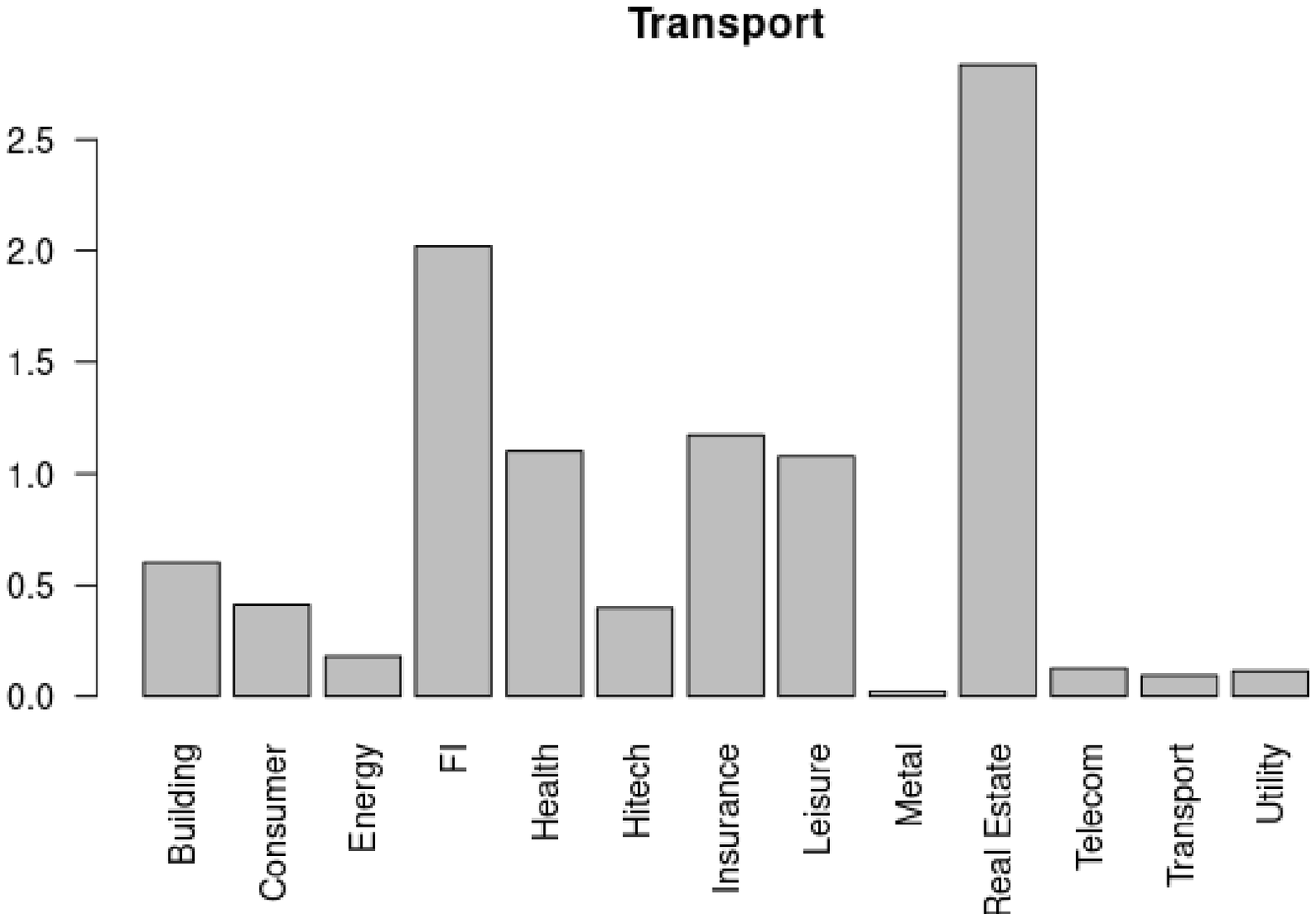} \\
\includegraphics[width=5.5cm]{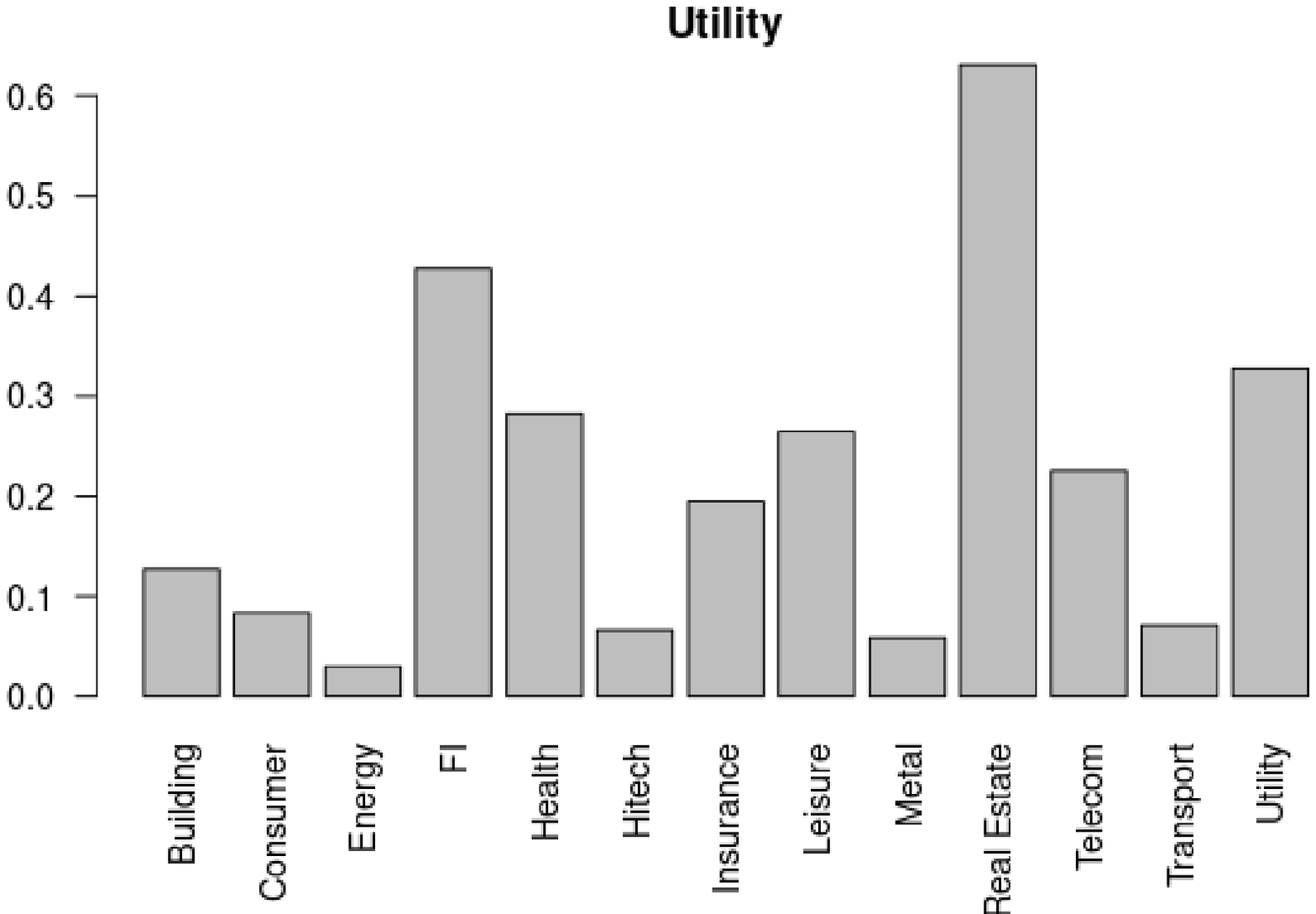}\\
\end{tabular}  
\caption{Impact of each sector $\mathbf{v}_{\infty}^{(i)}$ obtained using (1) MD-SE-NBD.
}
\label{impact1}
\end{figure}
\begin{figure}[htbp]
\begin{tabular}{ccc}    
\includegraphics[width=5.5cm]{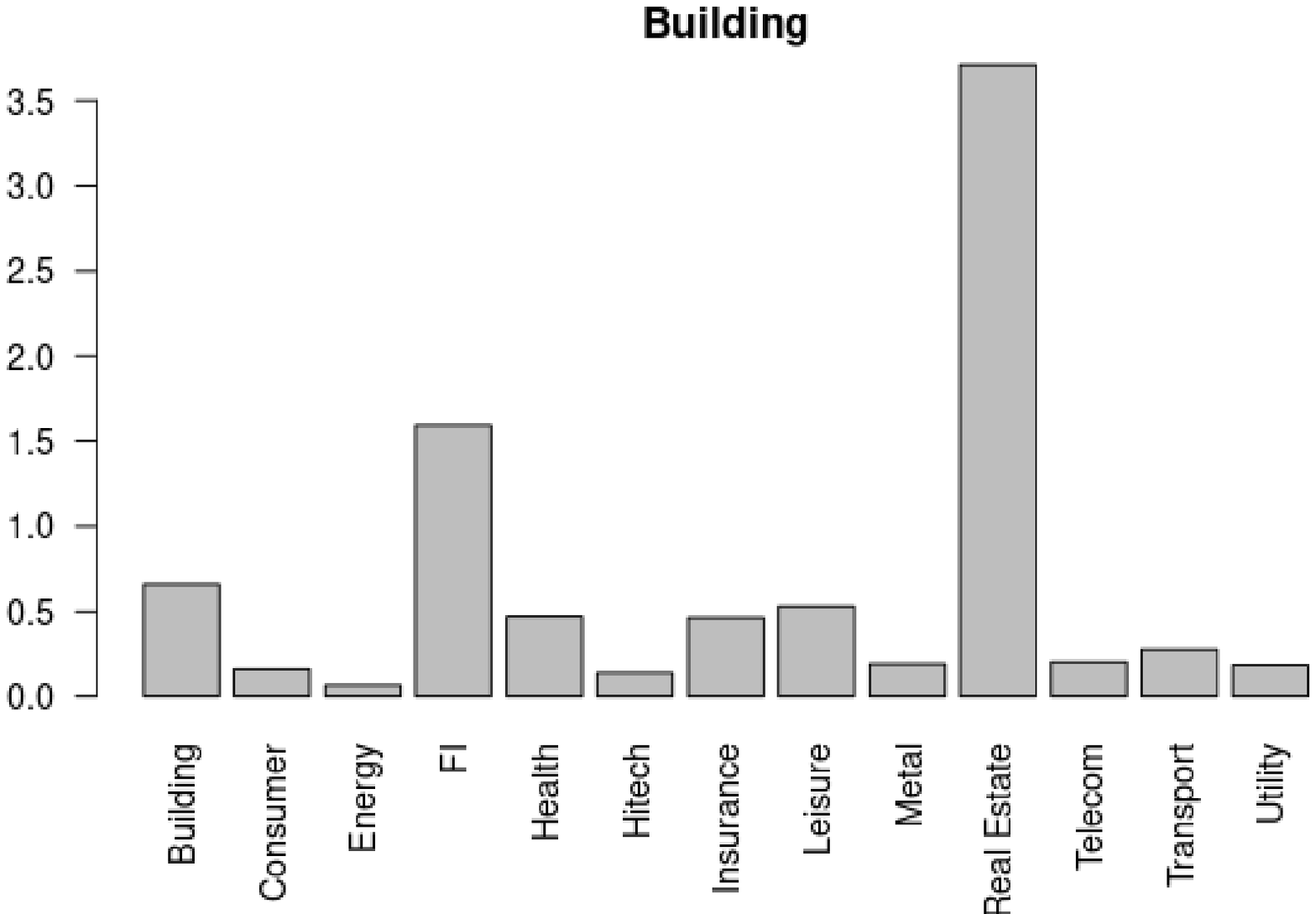} &
\includegraphics[width=5.5cm]{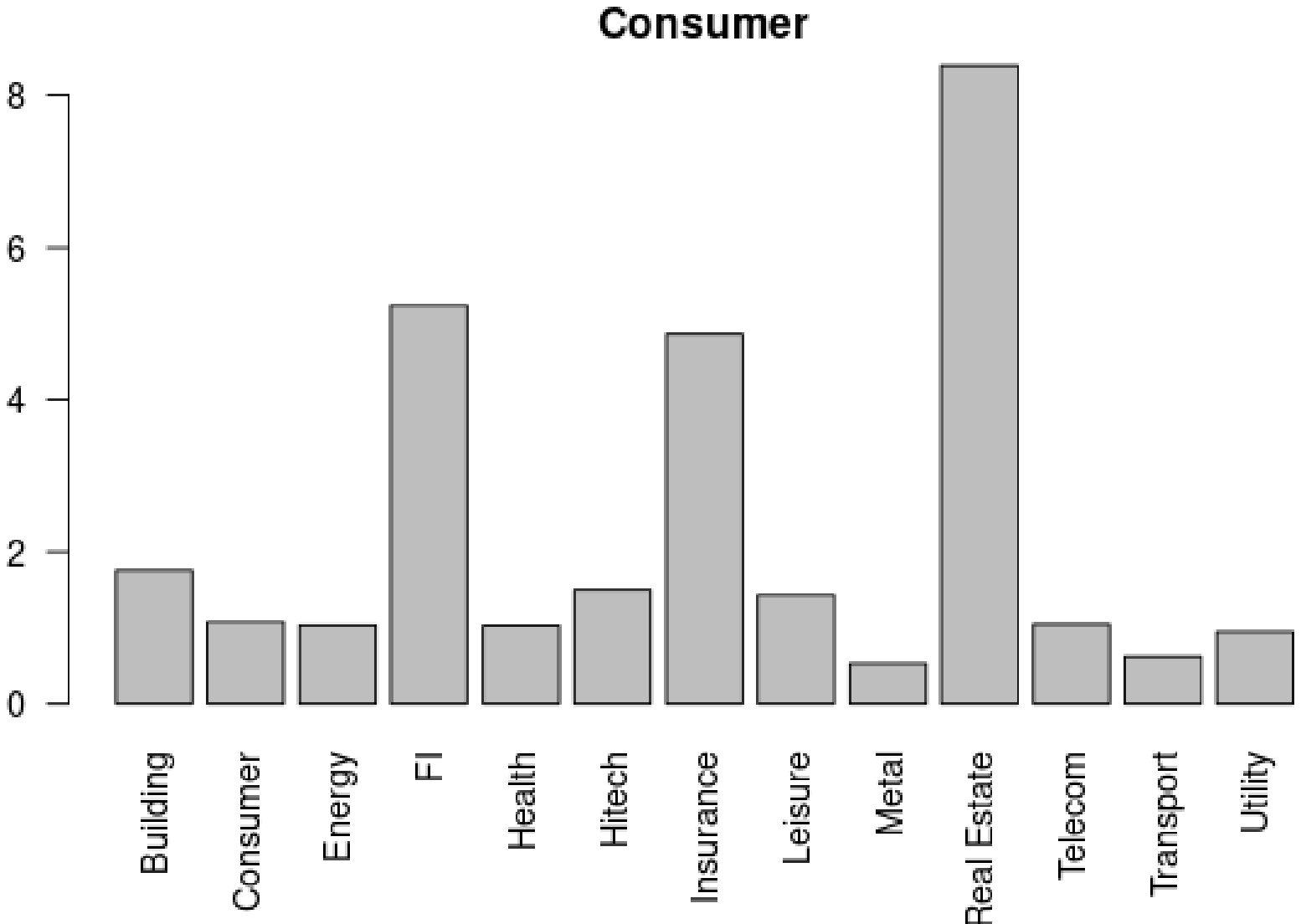} &
\includegraphics[width=5.5cm]{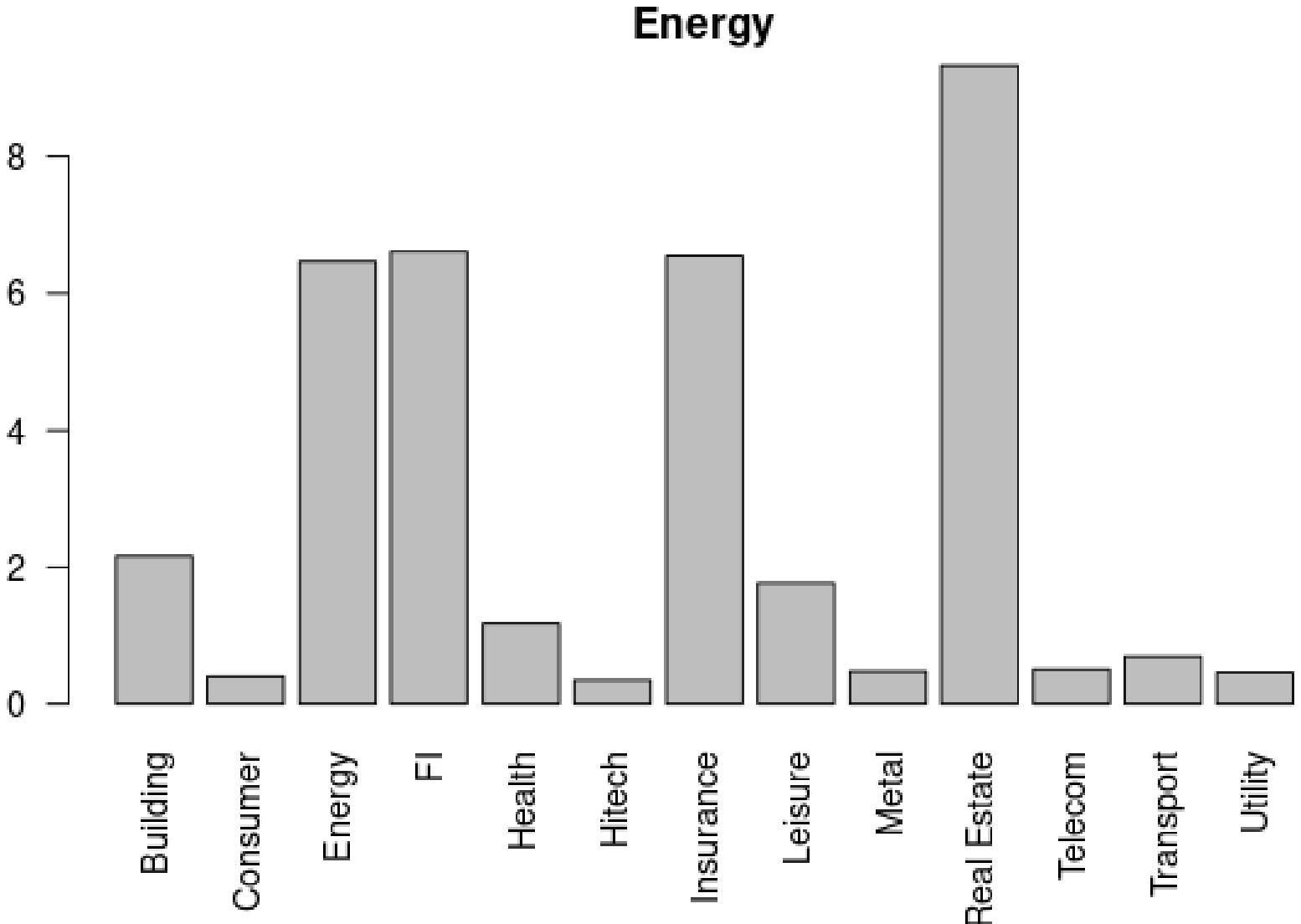} \\
\includegraphics[width=5.5cm]{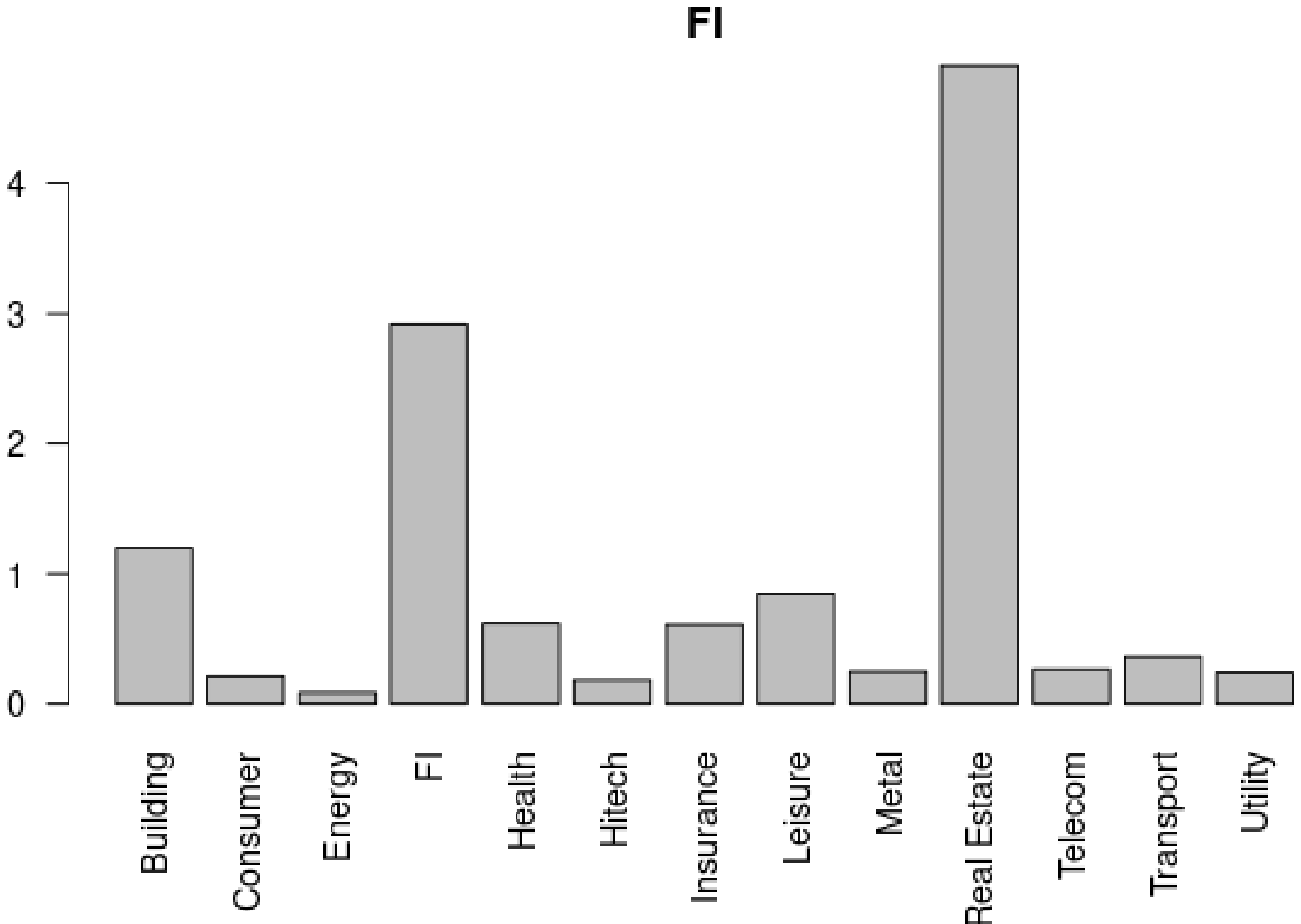} &
\includegraphics[width=5.5cm]{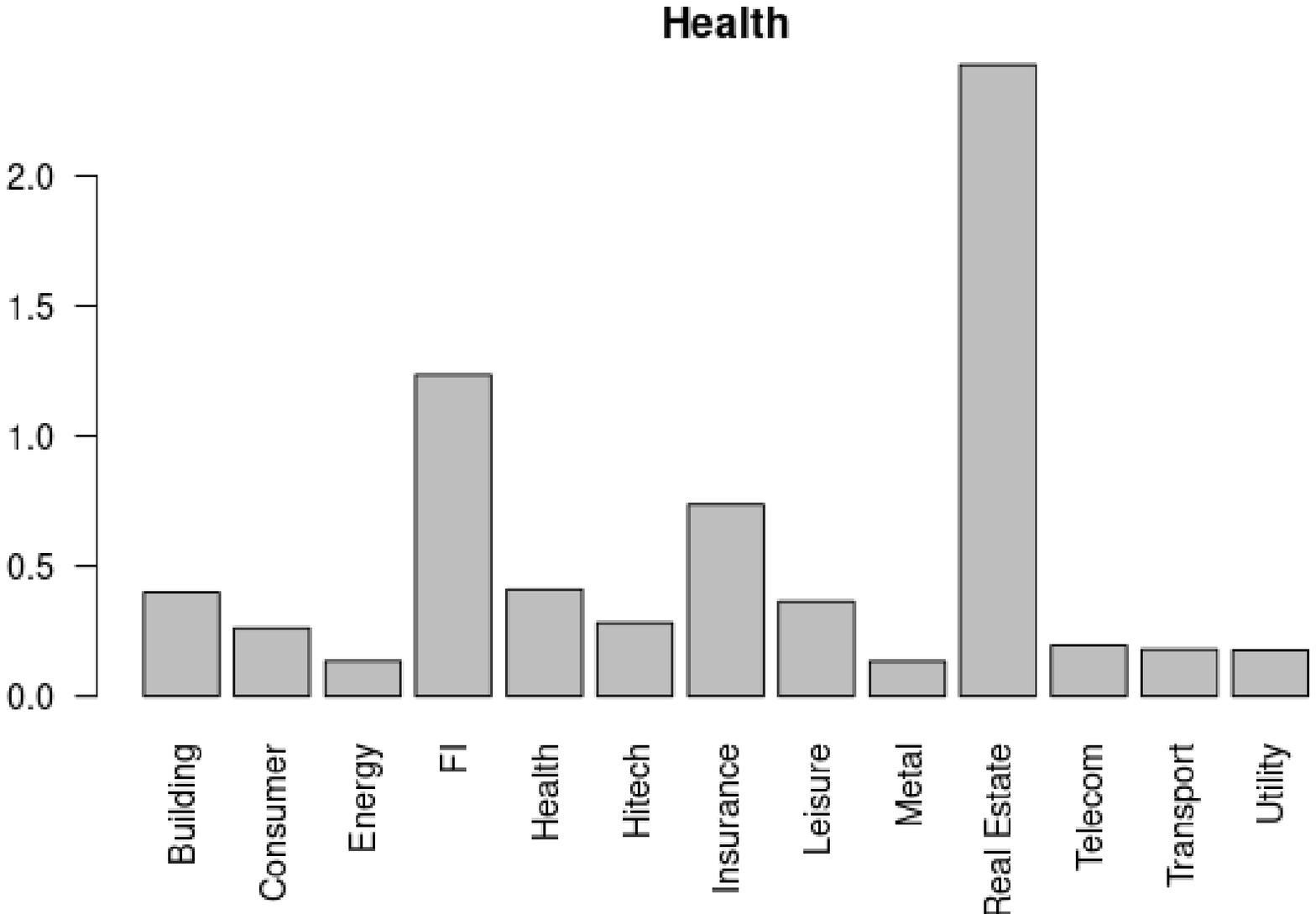} &
\includegraphics[width=5.5cm]{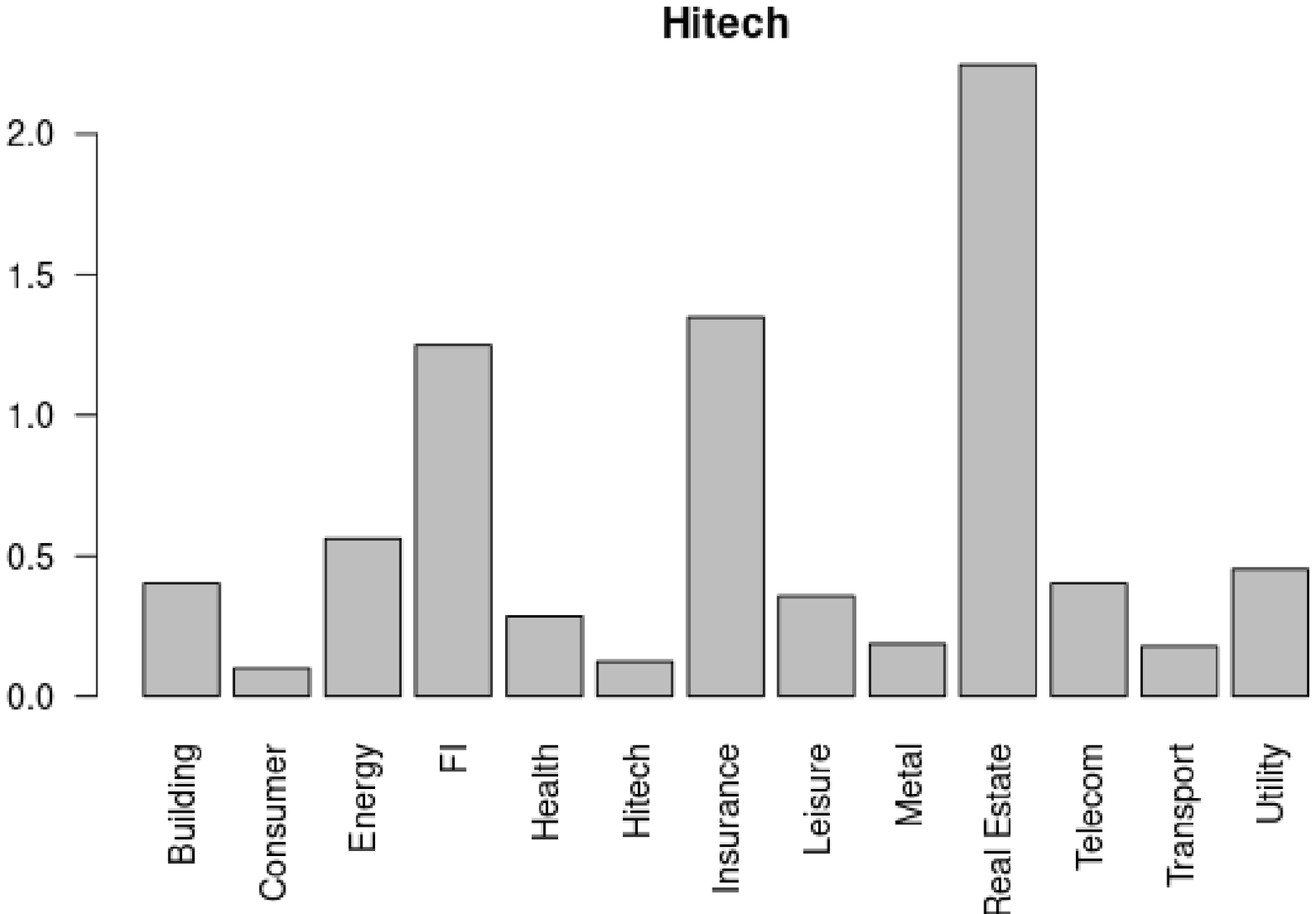} \\
\includegraphics[width=5.5cm]{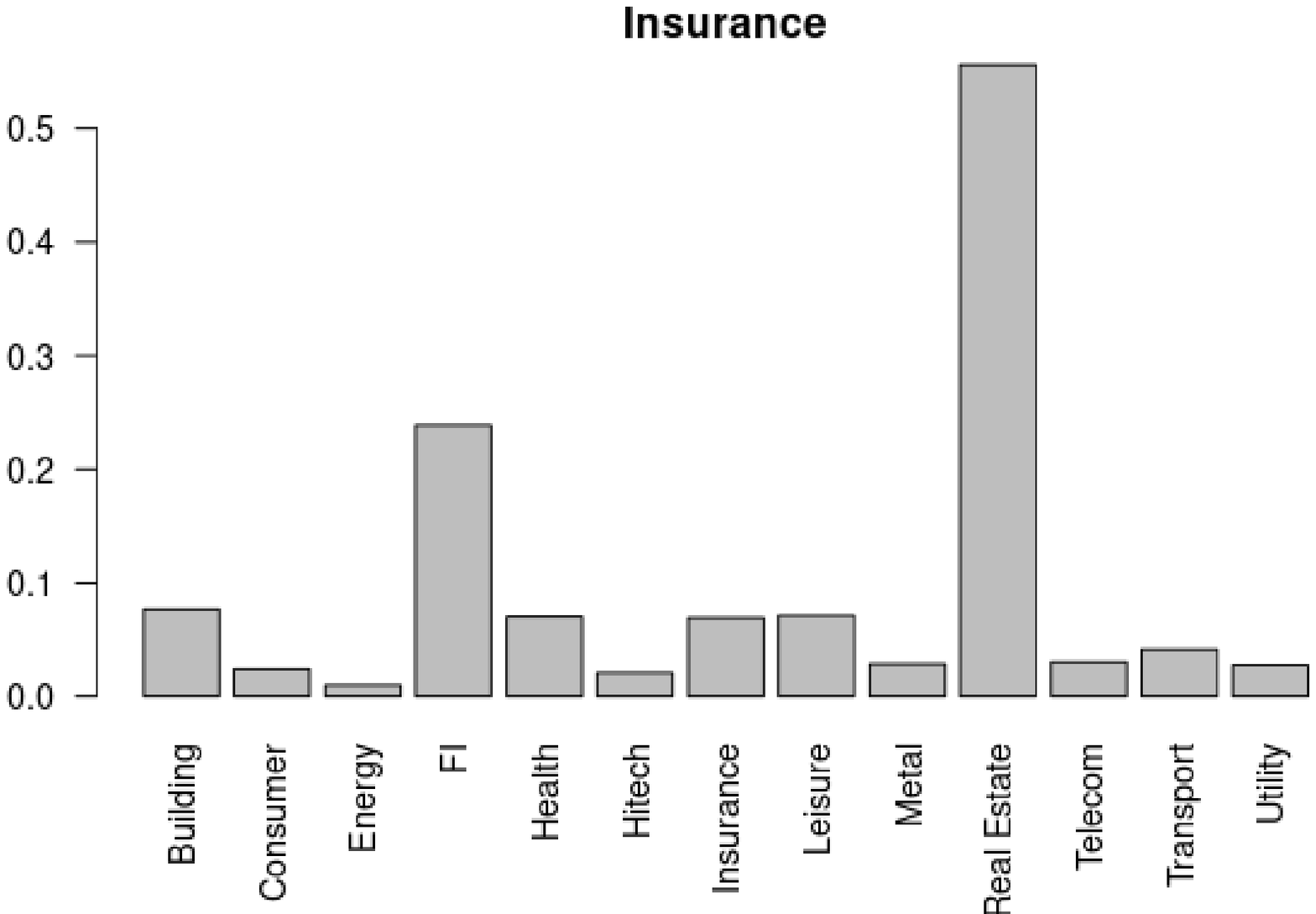} &
\includegraphics[width=5.5cm]{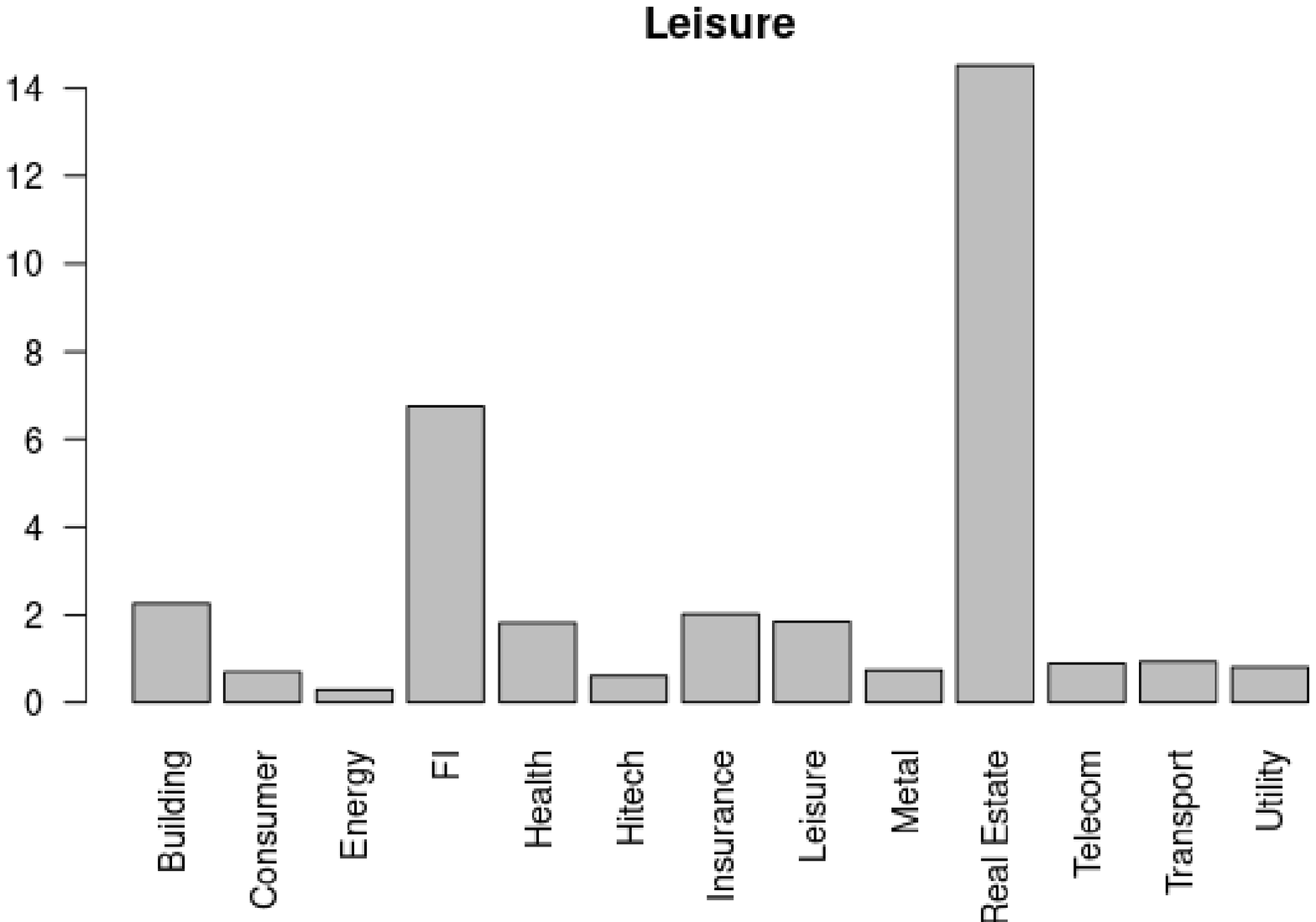} &
\includegraphics[width=5.5cm]{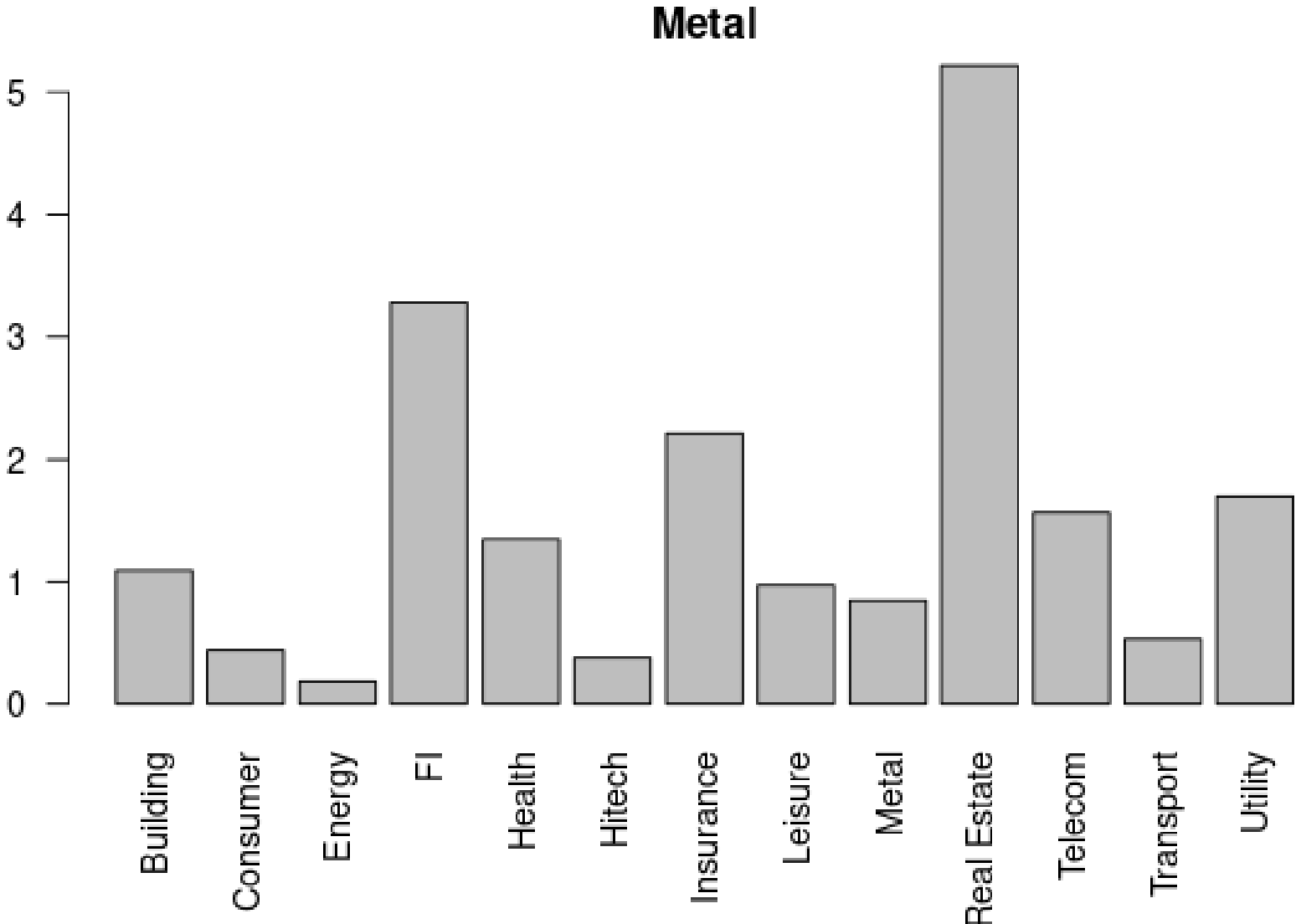} \\
\includegraphics[width=5.5cm]{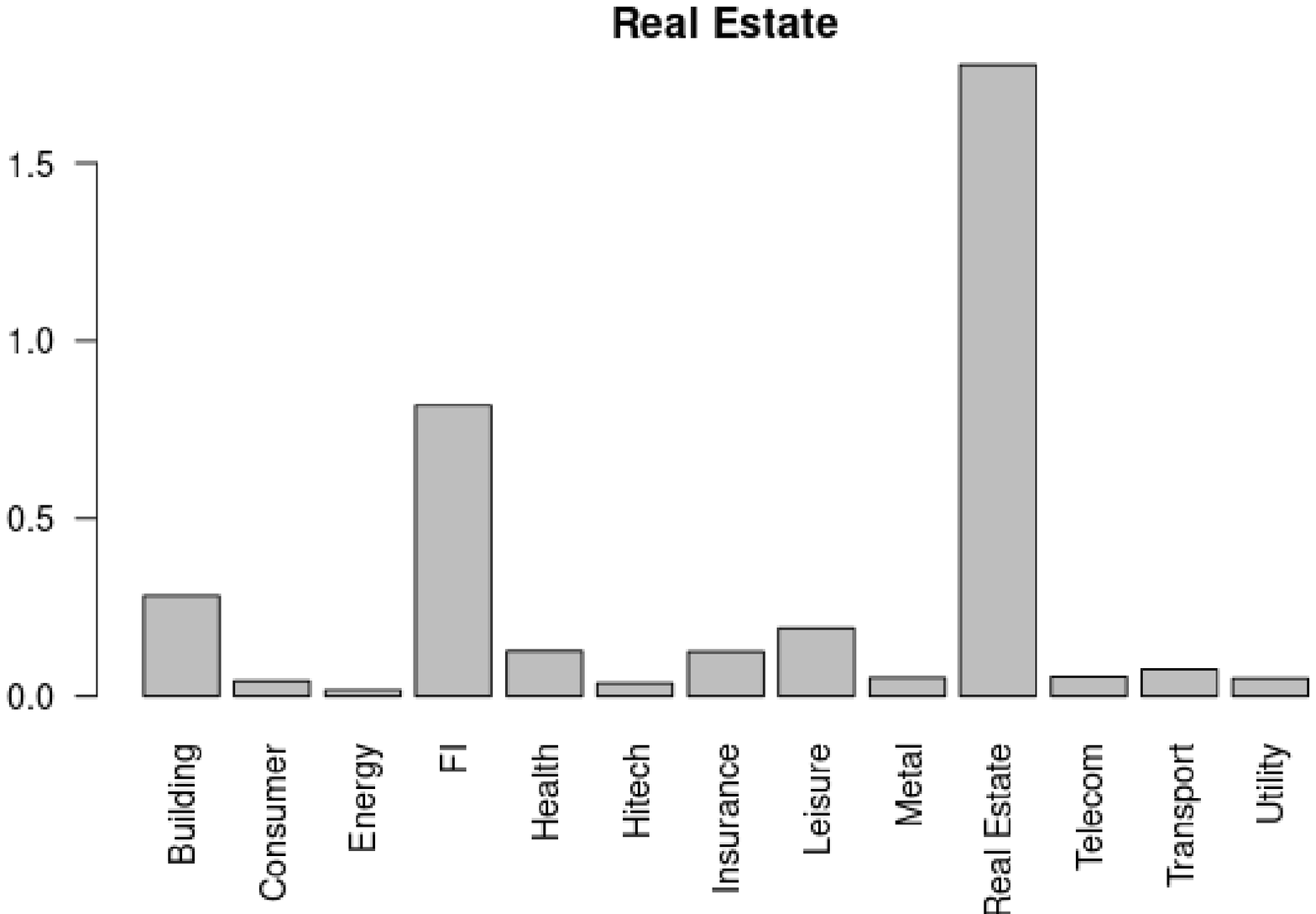} &
\includegraphics[width=5.5cm]{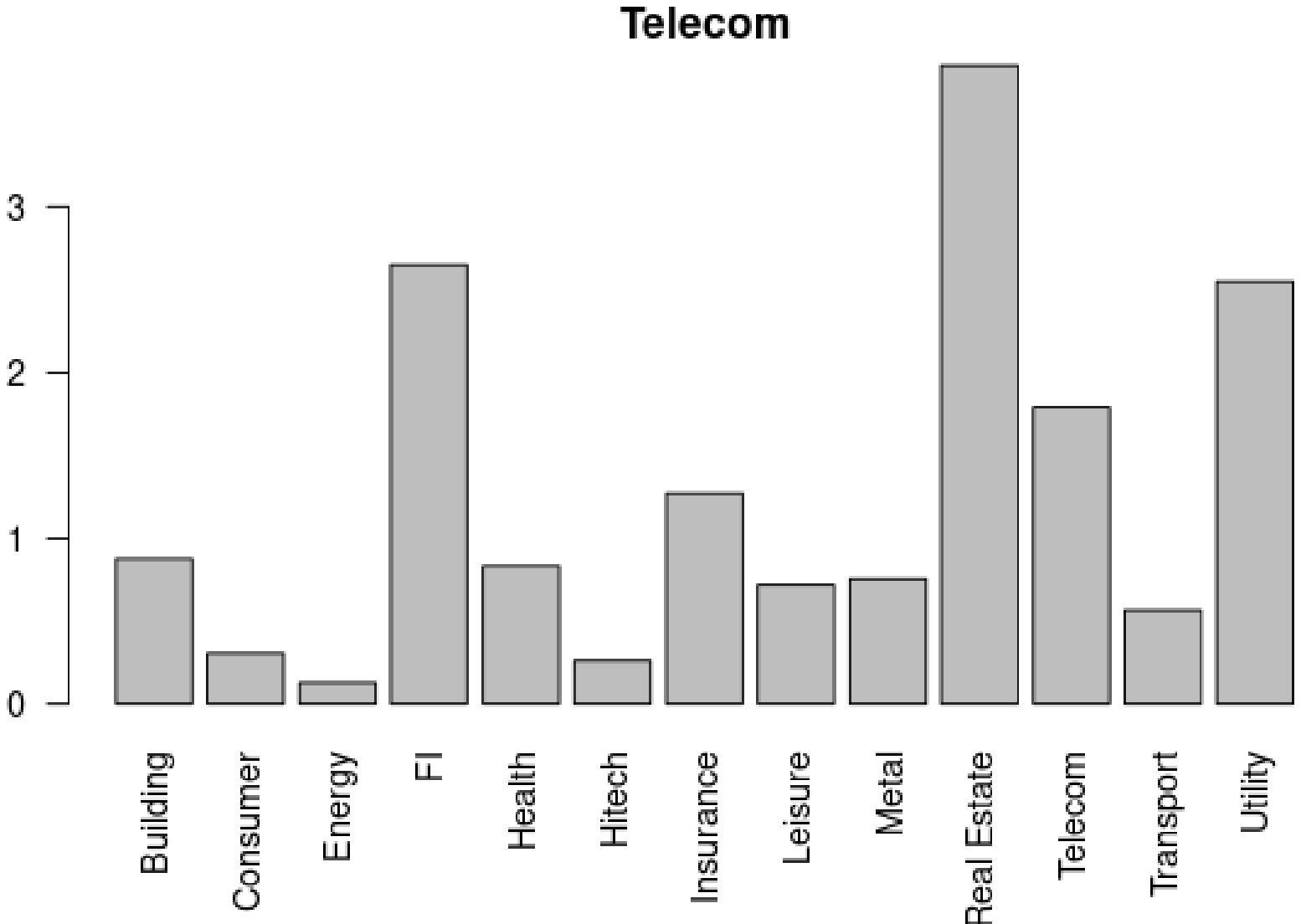} &
\includegraphics[width=5.5cm]{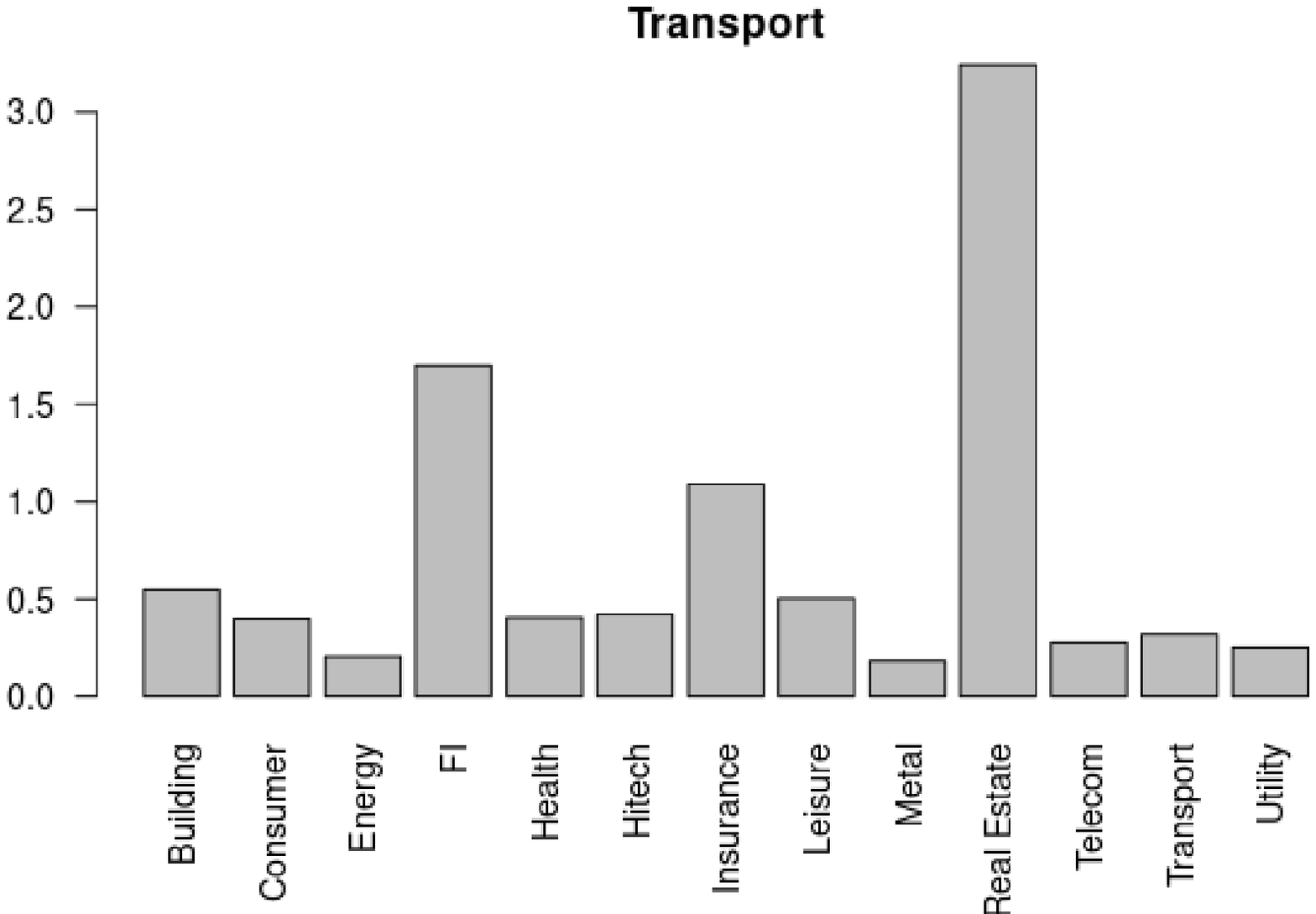} \\
\includegraphics[width=5.5cm]{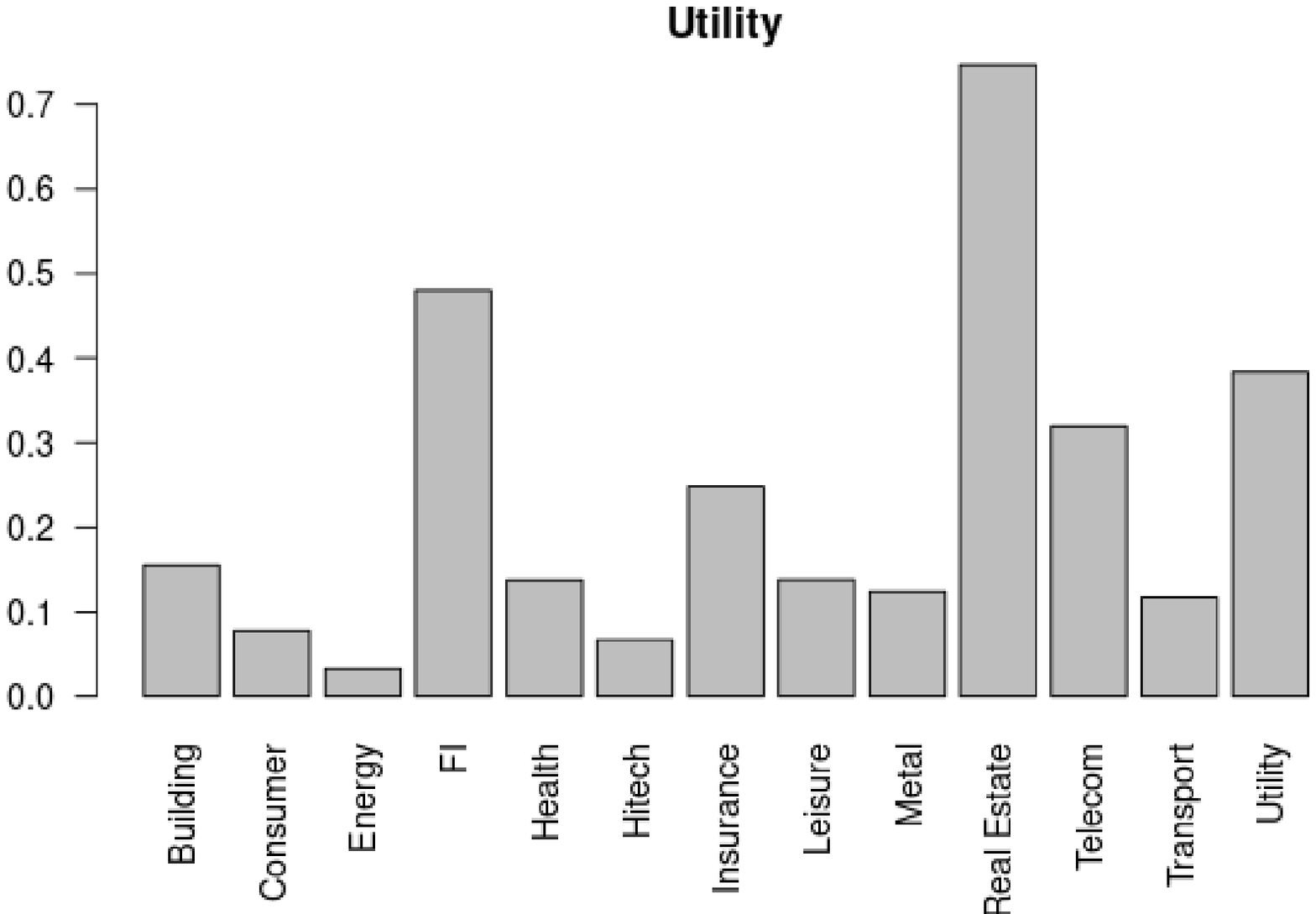}\\
\end{tabular}  
\caption{Impact of each sector $\mathbf{v}_{\infty}^{(i)}$ obtained using (2) MD-Hawkes.
}
\label{impact2}
\end{figure}
\begin{figure}[h]
\begin{center}
\begin{tabular}{c}
\begin{minipage}{0.5\hsize}
\begin{center}
\includegraphics[clip, width=8cm]{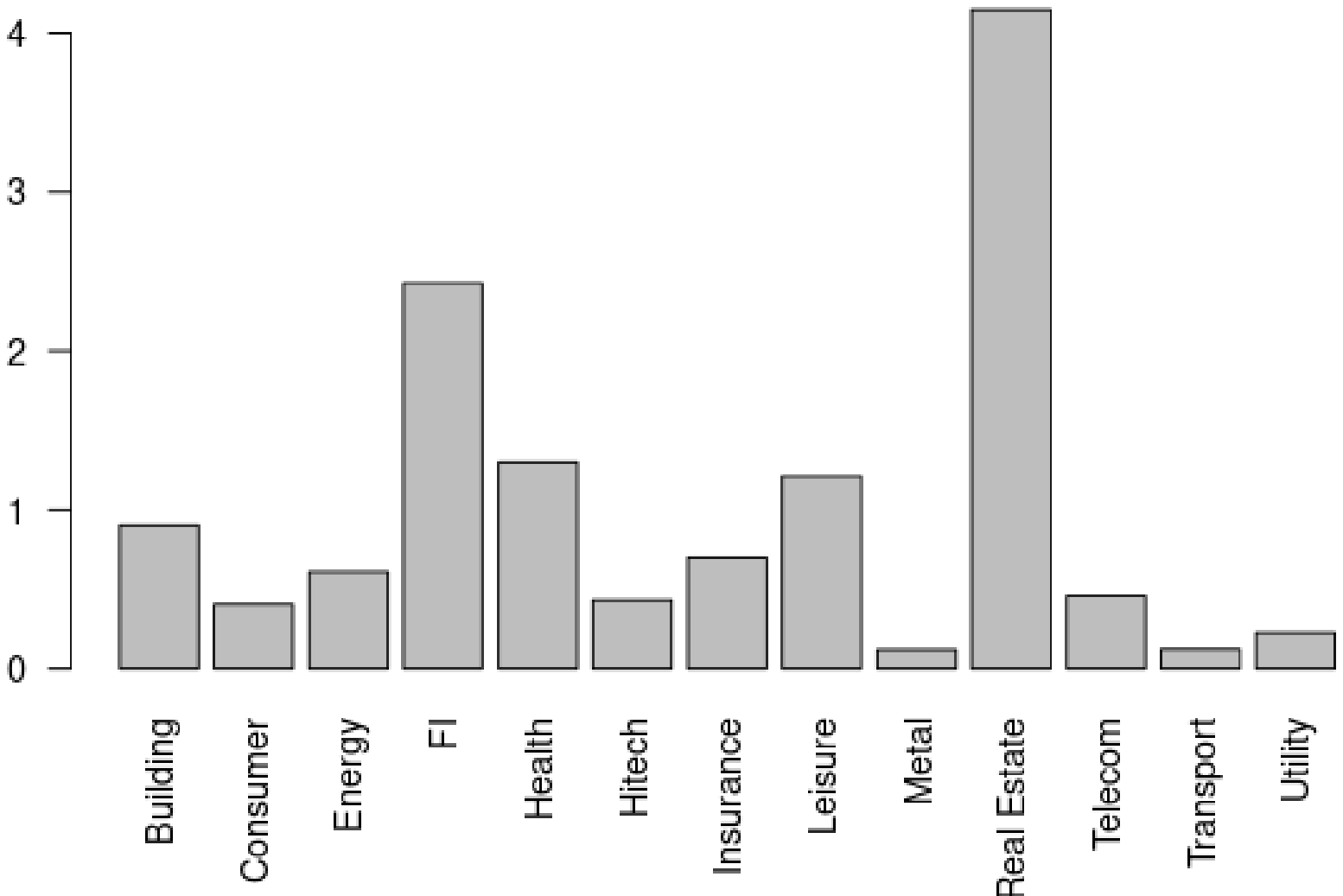}
\hspace{1.6cm} (a)
\end{center}
\end{minipage}
\begin{minipage}{0.5\hsize}
\begin{center}
\includegraphics[clip, width=8cm]{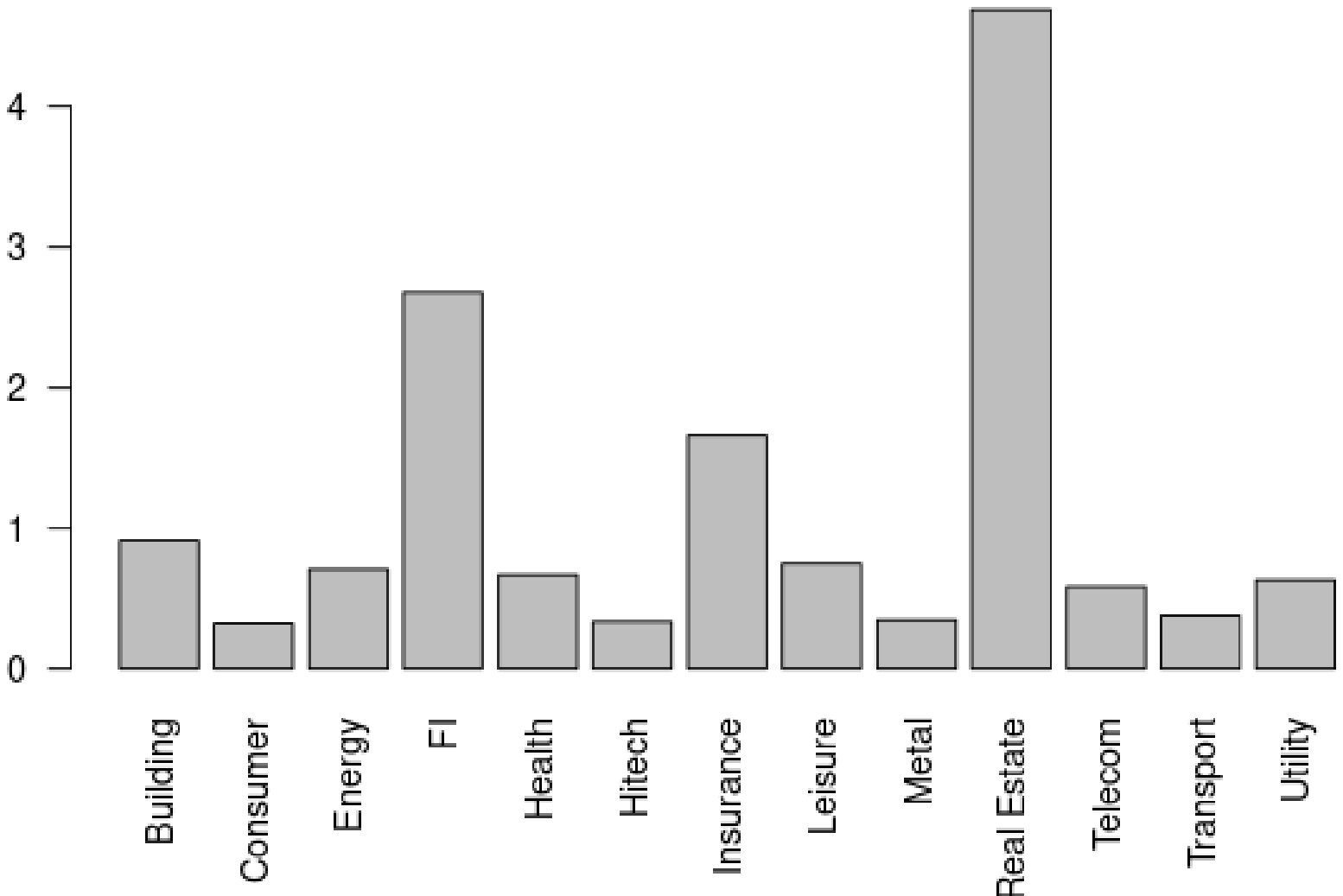}
 \hspace{1.6cm} (b)
\end{center}
\end{minipage}
 \end{tabular}
\caption{Impact $\mathbf{v}^{(i)}$ of the (a) MD-SE-NBD and (b) MD-Hawkes models}
\label{impact}
\end{center}
\end{figure}
\begin{figure}[h]
\includegraphics[width=110mm]{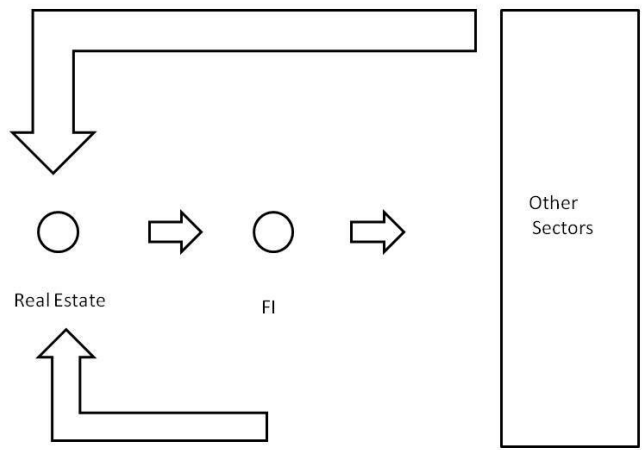}
\caption{Image of the contagion of shock among the sectors. It plays the role of amplifier of the shock.}
\label{cycle}
\end{figure}

To define the upstream and downstream sectors, 
we calculate the impacts $\mathbf{v}_{\infty}^{(i)}$
for (1) MD-SE-NBD and (2) MD-Hawkes, depicted
in Fig.\ref{impact1} and Fig.\ref{impact2}, respectively.
The figures \red{show} the contagion of a shock in each sector. 
The impact $\mathbf{v}_{\infty}^{(i)}$ denotes the
total increase in defaults due to the shock in the $i$-th sector. 
 We show $\mathbf{v}^{(i)}$ for (1) MD-SE-NBD and (2) MD-Hawkes
in Figs.\ref{impact} (a) and (b).
The real-estate and FI sectors were upstream in both these models.
If we apply the hybrid model, we can obtain almost the same result as with the MD-SE-NBD model.
Note that in the multiple sectors, the most affected sector is the real-estate sector, shown in Fig.\ref{impact1} and Fig.\ref{impact2}.
The shock in the real-estate sector affects the FI sector.
Through the FI sector, the shock spreads to all the sectors and
returns to the real-estate sector.
This is the cycle between the real-estate, FI, and other sectors. The cycle plays the role of an amplifier of the shock, depicted in Fig.\ref{cycle}.
This is consistent with the narrative of the Great Recession, 2009, because the origin is the shock of the price decrease in the
real-estate sector in the US.

\newpage
\begin{table}[tbh]
\caption{AIC for the discrete SE-NBD, discrete Hawkes, discrete  NBD, and discrete Hawkes processes }
\begin{center}
\begin{tabular}{|l|l|c|c|c|c|c|}
\multicolumn{4}{c}{}\\ \hline
&   &(1) MD-SE-NBD & (2)  MD-Hawkes  & (3)  SE-NBD& (4) Hawkes & (5) NBD \\
No.& Model & 
 AIC & AIC &AIC&AIC&AIC\\
 \hline \hline
1&Building& 369.76& 377.56 &386.65&403.04&420.21\\ \hline
2& Consumer& 616.63 & 626.16 &627.20&647.18&710.24\\ \hline
3& Energy & 498.82 & 552.08 &504.77&570.07&625.28\\ \hline
4& Financial Institution& 402.54& 408.21 &414.26&427.76&476.45\\ \hline
5&Health & 387.40 & 390.22 &389.28&395.27&420.75\\ \hline
6&Hi-tech&278.24 & 276.42 &290.52&289.49&311.62\\ \hline
7&Insurance & 253.37& 261.49&256.48 &267.26&257.72\\ \hline
8&Leisure & 543.29& 570.96&559.42 &608.11&625.08\\ \hline
9&Metal& 556.09& 582.94 &566.31 &609.69&621.96\\ \hline
10&Real Estate& 124.39 & 122.34  &133.37& 131.37&166.12\\ \hline
11& Telecommunication &327.39 & 331.41 &345.65& 359.99&426.96\\ \hline
12&Transport&359.57& 357.73& 378.34&377.81&405.68\\ \hline
13&Utility& 252.26& 266.78 & 270.09&313.79&285.89\\ \hline
\end{tabular}
\label{AIC}
\end{center}
\end{table}

\newpage
\begin{table}[tbh]
\caption{Parameter estimation for (1) discrete MD-SE-NBD}
\begin{center}
\begin{tabular}{|l|l|c|c|c|c|c|c|c|}
\multicolumn{4}{c}{}\\ \hline
No.& Model & 
 $K_0$ & $M_0/K_0$ &$S_{ij}$&&&&\\
 \hline \hline
1&Building& 0.41& 0.82 &$S_{11}=0.21$&$S_{18}=0.25$&&&\\ \hline
2& Consumer& 1.12 & 0.78 &$S_{22}=0.51$&$S_{24}=0.29$&$S_{26}=0.77$&$S_{27}=0.41$&\\ \hline
3& Energy & 0.00 & 0.00 &$S_{33}=0.86$&$S_{34}=0.17$&$S_{37}=0.51$&&\\ \hline
4& Financial Institution& 0.00& 0.38 &$S_{44}=0.53$&$S_{41}=0.28$&$S_{48}=0.14$&&\\ \hline
5&Health & 0.48 & 0.04 &$S_{55}=0.17$&$S_{52}=0.10$&$S_{58}=0.14$&&\\ \hline
6&Hi-tech&1.75 & 0.05&$S_{66}=0.12$&$S_{63}=0.04$&$S_{67}=0.29$&$S_{610}=0.19$&$S_{611}=0.12$\\ \hline
7&Insurance & 0.30& 0.47&$S_{77}=0.00$ &$S_{78}=0.11$&&&\\ \hline
8&Leisure & 0.35& 0.82&$S_{88}=0.20$ &$S_{84}=0.39$&$S_{85}=0.76$&$S_{810}=1.38$&\\ \hline
9&Metal& 0.74& 0.78& $S_{99}=0.30$&$S_{95}=0.41$&$S_{911}=0.48$&&\\ \hline
10&Real Estate& 23.84 & 0.00 &$S_{1010}=0.54$& $S_{104}=0.08$&&&\\ \hline
11& Telecommunication &0.00 & 0.38 &$S_{1111}=0.44$& $S_{118}=0.07$&$S_{119}=0.12$&$S_{1113}=0.39$&\\ \hline
12&Transport&4.45& 0.04&$S_{1212}=0.25$ &$S_{122}=0.09$&$S_{128}=0.09$&&\\ \hline
13&Utility& 0.02& 0.73 & $S_{1313}=0.07$&$S_{1311}=0.22$&$S_{1312}=0.24$&&\\ \hline
\end{tabular}
\label{MDSENBD}
\end{center}
\end{table}

\newpage

\begin{table}[tbh]
\caption{Parameter estimation for (2) discrete MD-Hawkes}
\begin{center}
\begin{tabular}{|l|l|c|c|c|c|c|c|c|}
\multicolumn{4}{c}{}\\ \hline
No.& Model & 
 $M_0$ & $S_{ij}$ &&&&&\\
 \hline \hline
1&Building& 0.16 &$S_{11}=0.24$&$S_{18}=0.25$&&&&\\ \hline
2& Consumer& 0.45 &$S_{22}=0.53$&$S_{24}=0.27$&$S_{26}=0.85$&$S_{27}=0.42$&&\\ \hline
3& Energy & 0.00 &$S_{33}=0.86$&$S_{34}=0.17$&$S_{37}=0.39$&&&\\ \hline
4& Financial Institution& 0.00 &$S_{44}=0.53$&$S_{41}=0.28$&$S_{48}=0.14$&&&\\ \hline
5&Health & 0.12  &$S_{55}=0.17$&$S_{52}=0.12$&$S_{58}=0.14$&&&\\ \hline
6&Hi-tech&0.09 &$S_{66}=0.12$&$S_{63}=0.04$&$S_{67}=0.29$&$S_{610}=0.19$&$S_{611}=0.12$&\\ \hline
7&Insurance & 0.14&$S_{77}=0.00$ &$S_{78}=0.11$&&&&\\ \hline
8&Leisure & 0.32&$S_{88}=0.25$ &$S_{84}=0.24$&$S_{85}=0.24$&$S_{89}=0.10$&$S_{810}=1.76$&$S_{812}=0.27$\\ \hline
9&Metal& 0.36& $S_{99}=0.30$&$S_{94}=0.14$&$S_{95}=0.37$&$S_{97}=0.37$&$S_{911}=0.45$&\\ \hline
10&Real Estate& 0.00 &$S_{1010}=0.54$& $S_{104}=0.08$&&&&\\ \hline
11& Telecommunication &0.00  &$S_{1111}=0.46$& $S_{114}=0.09$&$S_{119}=0.13$&$S_{1113}=0.40$&&\\ \hline
12&Transport& 0.19&$S_{1212}=0.25$ &$S_{122}=0.09$&$S_{128}=0.09$&&&\\ \hline
13&Utility& 0.00& $S_{1313}=0.00$&$S_{1311}=0.29$&$S_{1312}=0.22$&&&\\ \hline
\end{tabular}
\label{MDH}
\end{center}
\end{table}

\newpage

\begin{table}[tbh]
\caption{MAP estimation of the parameters for (3) discrete SE-NBD, (4) discrete Hawkes, and (5) NBD processes}
\begin{center}
\begin{tabular}{|c|l|lll|ll|ll|}
\multicolumn{2}{c}{}\\ \hline
&  & (3) SE-NBD  & && (4) Hawkes &  & (5) NBD  &\\
No.& Model & $K_0$&$M_0/K_0$& $S_{ii}$ & $M_0$&$S_{ii}$ & $K_0$ & $M_0/K_0$ 
\\ 
\hline \hline
1&Building& 0.58&0.65& 0.56 & 0.33&
0.62& 0.69& 0.79\\ \hline
2&Consumer& 1.06&0.66& 0.75 & 0.62&0.78& 1.33 & 0.49 \\ \hline
3& Energy&0.22 &1.40 & 0.88&0.27&0.90& 0.38 & 16 \\ \hline
4&Financial Institution& 0.37&0.56& 0.82 & 0.22&0.81& 0.78 & 0.72\\ \hline
5&Health& 0.48&0.41 & 0.81& 0.17&0.83& 0.98& 1.17 \\ \hline
6&Hi-tech &1.63&0.13 & 0.57 & 0.21&0.57& 1.32 & 2.70  \\ \hline
7&Insurance& 0.34&0.52 & 0.53 & 0.23&0.37& 0.67 & 1.86  \\ \hline
8&Leisure&0.41 & 1.11 & 0.78 & 0.43&0.79 &0.87 & 0.43\\ \hline
9& Metal& 0.62&0.99& 0.69& 0.55&0.73& 0.89 & 0.45 \\ \hline
10&Real Estate&2213709.45 &0.00 & 0.78 & 0.05&0.78& 0.48 & 2.70 \\ \hline
11&Telecommunication& 0.28&0.59 & 0.84 & 0.17&0.83& 0.34 & 0.35 \\ \hline
12&Transport& 2.84&0.15 & 0.49 & 0.42&0.48 &2.22 & 2.70 \\ \hline
13&Utility&0.19&1.07& 0.56& 0.16&0.65 &0.31& 0.69 \\ \hline
\end{tabular}
\label{Fig4}
\end{center}
\end{table}
\newpage

\newpage

\newpage
\newpage

\section{VI. Concluding Remarks}

In this study, we applied a multidimensional
self-exciting negative binomial distribution (MD-SE-NBD) process
 to default portfolios with 13 sectors.
 The SE-NBD process is a Poisson process with a gamma-distributed intensity function.
 Using the multidimensional SE-NBD process,
 we could estimate the interactions between  these sectors
 as a network.
 We compared these results to those obtained with the multidimensional Hawkes (MD-Hawkes) process, which 
  is a Poisson process with a zero-variance intensity function.
  By applying the empirical data, we created networks of
  contagious defaults among the sectors.
We confirmed that the upstream of contagious defaults is
the real-estate and financial-institution (FI) sectors.
These conclusions are consistent with our assessment of the default contagions.

\appendix
\section{Appendix A. Proof of the impact calculation}
\red{
In this section, we present the calculation of 
Eq.(\ref{im}).
We consider an event at $t=0$ and the number of affected events.
We consider the sequence $a_t$ that is the effect at $t$ and $v_t=\sum_t a_t$.
The sequence is 
 $a_0=1$ and 
 \begin{equation}
a_{t+1}=\frac{M_0}{L_0}\sum_{s=0}^{t}\hat{d}_{t-s} a_s=
\frac{M_0}{L_0}\sum_{s=0}^{t}r^{t-s} a_s.
 \end{equation}
After solving this recurrence formula, we obtain
\begin{equation}
 a_{t+1}=\frac{M_0}{L_0}(\frac{M_0}{L_0}+r)^t.   
\end{equation}
 Subsequently, the impact is,
\begin{equation}
 v_t=\frac{M_0}{L_0}\frac{1-(r+\frac{M_0}{L_0})^t}{1-r-\frac{M_0}{L_0}}.
 \end{equation}
In the limit $t\rightarrow \infty$, we can obtain Eq.(\ref{im}).
}


\section{Appendix B.  Impact and Branching process}

In this section, we consider the relationships between 
impact and branching process to clarify the mean of the impact analysis based on the single-line case.
We consider the case of
$\hat{d}_1=1$ and $\hat{d}_s=0$, $s=2,3,\cdots$ for Eq.(\ref{SENBD}) and Eq.(\ref{Hawkes}), which denote the
 discrete NBD and Hawkes processes, respectively.
In this case, we obtain $\hat{T}=1$.
This corresponds to an event that affects only the next term.

In Eq.(\ref{SENBD}) and Eq.(\ref{Hawkes}), $M_0$ denotes the birth probability of the 0-th generation, which corresponds to the parent events, and the second term corresponds to the birth of the offspring events.
 
 We focus on a single parent event and its offspring events.
 $Y_s$ denotes the number of offspring events.
 The parent event occurs at $t$ and the offspring events occur at the $s$-th generation at $t+s$.
Here, we set the parent event to the 0-th generation.

\begin{figure}[h]
\includegraphics[width=110mm]{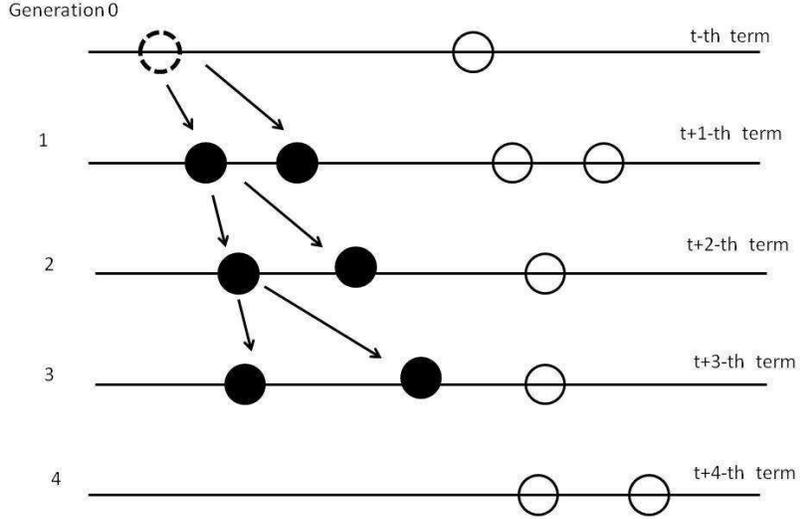}
\caption{Parent and offspring events. We focus on a single parent event and its offspring events. The dotted circle denotes the parent event, while the black and white circles denote the offspring and other events, respectively.}
\label{BP}
\end{figure}

We define $p_k$ as the offspring distribution for $k$ children.
With probability $p_0$, there are no offspring events.
Using Eq.(\ref{SENBD}) and Eq.(\ref{Hawkes}), the offspring distribution becomes $\mbox{NBD}(K_0/L_0,M_0/K_0 L_0)$, while the Poisson distribution becomes $\mbox{Poisson}(M_0/L_0)$.
Because these distributions have a reproductive function, the sum of the distributions becomes Eq.(\ref{SENBD}) and Eq.(\ref{Hawkes}).
This corresponds to the decomposition of the SE-NBD and Hawkes processes, focusing on a single parent.
This is a discrete-time branching process.
An image of the process is shown in Fig. \ref{BP}.

In summary, the process has the following properties:
\begin{quote}
\begin{itemize}
\item $Y_0=1$ this is the parent event.
\item When $Y_n=i$, where $i=0,1,2,\cdots$, if $i=0$, $Y_{n+1}=0$. This corresponds to the extinction of the offspring event.
\item If $i>0$, new offspring events occur independently with the probability $\mbox{NBD}(K_0/L_0, M_0/K_0)$ or  $\mbox{Poisson}(M_0/L_0)$ at the next term. 
\end{itemize}
\end{quote}

Note that the impact is considered as the sum of the offspring events $\sum_k Y_k-1$. This impact does not include the parent event.

Here, we consider the extinction probability $\epsilon$ and probability of $Y_n=0$ for any $n$, considering the phase transition.
After this term, there are no offspring events for the parent event.
We can obtain the self-consistent equation  
 \begin{equation}
     x=\sum_{k=0}^{\infty}p_k x^k,
     \label{mean}
 \end{equation}
 and its solution is $x=\epsilon$.
 Here, we introduce the moment-generating function
 \begin{equation}
 f(x)=\sum_{k=0}^{\infty}p_k x^k.
 \end{equation}
We rewrite Eq.(\ref{mean}) using the generating function,
\begin{equation}
x=f(x).
\label{bp}
\end{equation}
Eq.(\ref{bp}) has the solution $x=1$ because $f(1)=\sum_k p(k)=1$.
If $f'(1)>1$, then there is only one solution, $x=1$; accordingly, the probability is 1.
If $f'(x)<1$, then there are two solutions, where the smaller solution denotes the extinction probability $\epsilon$.
The transition point is $f'(1)=1$.
This is known as the branching process phase transition. 
Note that the order parameter of the phase transition is the survival probability $1-\epsilon$.

We change the variable from $x$ to $y=1-x$ and rewrite
Eq.(\ref{bp}) around $y\sim 0$,
\[
1-y=f(1-y)=1-f'(1)y+f''(1)y^2/2-\cdots.
\]
Then, we can obtain the critical exponent $\beta=1$ at the transition point $f'(1)=M_0/L_0=1$,
\[
y\sim2(f'(1)-1)/f''(1),
\]
where $f''(1)>0$.

We consider the transition point from an impact analysis standpoint.
The impact is the expected value of the number of offspring events,  $v_{\infty}=\sum_{k=1}^{\infty} (M_0/L_0)^k=M_0/L_0\cdot1/(1-M_0/L_0)$ for  the SE-NBD and Hawkes processes determined using Eq.(\ref{im}), where we assume that $\hat{d}_1=1$ and $\hat{d}_s=0$, $s=2,3,\cdots$.
Hence, we can confirm that the transition point is $M_0/L_0=1$. 
At this limit, the survival probability is 1.

\section{Appendix C. Correlation functions of MD-SE-NBD process}
In this section, we consider the correlation functions for the MD-SE-MBD process.

The covariance density function is defined as 
\begin{equation}
    Cov[X_t^{(i)},X_{t'}^{(j)}]=E[X_t^{(i)},X_{t'}^{(j)}]-E[X_t^{(i)}]E[X_{t'}^{(j)}]
    =E[X_t^{(i)},X_{t'}^{(j)}]-\bar{v}_i\bar{v}_j\Delta^2=C_{ij}^{*}(t, t')\Delta^2,
    \label{21}
\end{equation}
where $C^{*}_{(ij)}(t, t')$ denotes the covariance  density function.

We can obtain for $t=t'$,
\begin{equation}
 E[X_t^{(i)},X_{t'}^{(i)}]=E[(X_t^{(i)})^2]=(1+\omega'_i)\bar{v}_i\Delta.   
\end{equation}
and 
\begin{equation}
 E[X_t^{(i)},X_{t'}^{(j)}]=\bar{v}_i\bar{v}_j\Delta^2,   
\end{equation}
because the processes in lines $i$ and $j$ at the same time are independent.
Note that the difference between the SE-NBD and
Hawkes processes is only the correlation at the same time $t=t'$.
Hence, we can decompose the covariance density function as
\begin{equation}
C^*_{(ij)}(t,t')=(1+\omega'_i)\bar{v}_i\delta_{ij}\delta(t-t')+C_{(ij)}(t,t'),  
\label{22}
\end{equation}
where $C_{ij}(t,t')$ is the  correlation function, 
$\delta_{ij}$ is the Kronecker’s delta, and $\delta(x)$ is the delta function.

If we set $\tau=t'-t$, we can rewrite Eq.(\ref{21}) using Eq.(\ref{22}),
\begin{equation}
    E[X_t^{(i)},X_{t+\tau}^{(j)}]=[C_{ij}(\tau)+\bar{v}_i\bar{v}_j+(1+\omega'_i)\bar{v}_i\delta_{ij}\delta(\tau)]\Delta^2.
\end{equation}

Here we define $g_{ij}(x)=\tilde{\omega}_{ij}\hat{d}_t^{(i)}(x)$ and set $\tau>0$.

We can calculate the following correlations:
\begin{eqnarray}
E[X_t^{(i)} X_{t+\tau}^{(k)}]&=&[C_{ik}(\tau)+\bar{v}_i\bar{v}_k]\Delta^2
\nonumber \\
&=&\theta_0^{(k)}\bar{v}_i\Delta^2+\sum_{j=1}^{D}\lim_{\Delta\rightarrow 0}\sum_{s=0}^{\infty}g_{kj}(s\Delta)E[X_t^{(i)} X_{t+\tau-(s+1)\Delta}^{(j)}]\Delta,
\nonumber \\
&=&\theta_0^{(k)}\bar{v}_i\Delta^2+\sum_{j=1}^{D}\lim_{\Delta\rightarrow 0}\sum_{s=0}^{\infty}g_{kj}(s\Delta)[C_{ij}(\tau-(s+1)\Delta)+\bar{v}_i\bar{v}_j+(\omega'_i+1)\bar{v}_i\delta_{ij}\delta(\tau-(s+1)\Delta)]\Delta^3,
\nonumber \\
&=&\theta_0^{(k)}\bar{v}_i\Delta^2+\Delta^2\sum_{j=1}^{D}\int_{0}^{\infty}g_{kj}(w)[C_{ij}(\tau-w)+\bar{v}_i\bar{v}_j+(\omega'_i+1)\bar{v}_i\delta_{ij}\delta(\tau-w)]dw,
\nonumber \\
&=&\theta_0^{(k)}\bar{v}_i\Delta^2+\Delta^2(\omega'_i+1)\bar{v}_i g_{ii}(\tau)+\Delta^2\sum_{j=1}^{D}\int_{0}^{\infty}g_{kj}(w)[C_{ij}(\tau-w)+\bar{v}_i \bar{v}_j)]dw,
\nonumber \\
&=&\bar{v}_i\bar{v}_k\Delta^2+\Delta^2(\omega'_i+1)\bar{v}_i g_{ii}(\tau)+\Delta^2\sum_{j=1}^{D}\int_{0}^{\infty}g_{kj}(w)C_{ij}(\tau-w)dw
\nonumber 
\end{eqnarray}
where we used the mean-field approximation defined in Eq.(\ref{M3}),
\begin{equation}
\bar{v}_k=\theta_0^{(k)}+ \sum_{j=1}^{D}\int_{0}^{\infty}g_{kj}(w)\bar{v}_j dw.   
\end{equation}
We can obtain the integral equation for $\tau>0$ as
\begin{equation}
    C_{ik}(\tau)=(\omega'_i+1)\bar{v}_i g_{ii}(\tau)+\sum_{j=1}^{D}\int_{0}^{\infty}
    g_{kj}(w)C_{ij}(\tau-w)dw.
    \label{ieq2}
\end{equation}
This is the integral equation for the correlation functions.
The difference between the SE-NBD and Hawkes processes
is only the first term, which signifies the effect of correlation at the same time.

We can rewrite Eq.(\ref{ieq2}) as
\begin{equation}
  C_{ik}(\tau)=(\omega'_i+1)\bar{v}_i g_{ii}(\tau)+\sum_{j=1}^{D}\int_{0}^{\tau}
    g_{kj}(w)C_{ij}(\tau-w)dw
    +\sum_{j=1}^{D}\int_{0}^{\infty}
    g_{kj}(\tau+w)C_{ji}(w)dw,
    \label{ieq3}
\end{equation}
where we use the relation $C_{ij}(\tau)=C_{ji}(-\tau)$.

In the single-line case 
Eq.(\ref{ieq3}) becomes
\begin{equation}
    C(\tau)=(\omega'+1)\bar{v}g(\tau)+\int_{0}^{\tau}
    g(w)C(\tau-w)dw+\int_{0}^{\infty}
    g(\tau+w)C(w)dw.
\end{equation}
Here, we set the kernel function $g(\tau)=ab\exp (-b\tau)$.
We can solve this equation using Laplace transformations,
\begin{equation}
    C(\tau)=\frac{ab(\omega'+1)\bar{v}}{2(1-a)}\exp [-b(1-a)\tau].
\end{equation}
Based on these results, the difference between the SE-NBD and Hawkes processes is only the intensity of the autocorrelation function.
In the case of $\omega'=0$, it corresponds to the correlation function of the Hawkes process.

\end{document}